\documentclass[11pt]{article}
\usepackage{jheppub}
\usepackage{bbold}
\usepackage{tikz}
\usepackage{etaremune}
\usepackage{enumitem}
\usepackage{hyperref}
\usetikzlibrary{arrows.meta}
\tikzstyle{bag} = [align=center]
\usetikzlibrary{decorations.pathmorphing}
\def\mfa{{\mathfrak{a}}}

\def\mfq{{\mathfrak{q}}}
\def\mfn{{\mathfrak{n}}}
\def\mfns{{\mbox{\footnotesize ${\mathfrak{n}}$}}}
\def\fq{q\hspace{0.25pt}}
\def\be{\begin{equation}}
\def\ee{\end{equation}}
\def\bea{\begin{eqnarray}}
\def\eea{\end{eqnarray}}
\def\cH{{\cal H}}
\def\J{{\cal J}}
\def\HH{{\bf H}}
\def\ra{\rangle}
\def\la{\langle}
\setlist[itemize]{leftmargin=6mm}
\setlist[enumerate]{leftmargin=6mm}
\usetikzlibrary{decorations.pathmorphing}
\tikzset{snake it/.style={decorate, decoration=snake}}
\def\darkblue{blue!85!black}
\def\darkred{red!60!black}
\def\darkgreen{green!50!black}

\newcommand{\Op}{\mathrm{Op}}
\newcommand\bC{\mathbb{C}}
\newcommand\bR{\mathbb{R}}
\newcommand\bZ{\mathbb{Z}}
\newcommand\cM{\mathcal{M}}
\newcommand\tH{T}
\def\yp{x}
\def\ym{y}
\def\cA{{\cal A}}
\def\Tr{{\rm Tr}}
\title{SYK-Schur duality: Double scaled SYK correlators from  ${\cal N}=2$ supersymmetric gauge theory}
\author[1]{Davide Gaiotto,}
\author[2,3]{Herman Verlinde}

\affiliation[1]{Perimeter Institute for Theoretical Physics, 31 Caroline Street North, Waterloo, ON N2L 2Y5, Canada}
\affiliation[2]{Department of Physics,  Princeton University,
Princeton, NJ 08544, USA}
\affiliation[3]{School of Natural Sciences, Institute of Advaced Study,  Princeton, NJ 08540, USA}
\date{}
\abstract{We propose a triality relating the Double-Scaled SYK model, $SL(2,\bC)$ Chern-Simons theory on a disk with an irregular singularity at the center and the outcome of ``real Schur quantization'' applied to $SU(2)$ Seiberg-Witten theory with Neumann boundary conditions. We give supporting evidence for our conjecture by establishing a precise match between a general class of correlators in all three systems.  }
\begin{document}
\addtolength{\abovedisplayskip}{.5mm}
\addtolength{\belowdisplayskip}{.5mm}

\maketitle

\pagebreak

\section{Introduction and Summary}

\vspace{-1mm}

Quantum equivalences between different physical systems, such as gauge-gravity duality \cite{Maldacena:1997re} and the AGT correspondence \cite{Alday:2009aq}, provide powerful insights and computational tools for studying the physical properties of both systems. Frequently, the algebra of observables inherit remarkable algebraic or geometrical characteristics from the duality that are evident from one side but would seem obscure from the other side. With this motivation, we propose and investigate a new exact correspondence between the following exactly soluble but seemingly unrelated quantum mechanical systems:\\[2mm]
{(A}) The partition function and correlation functions of local operators in the double scaling limit of the Sachdev-Ye-Kitaev model \cite{kitaevTalks,Sachdev:1992fk,Maldacena:2016hyu,Berkooz:2018jqr} at infinite temperature. \\[2mm]
{(B}) A collection of ``Schur half-indices'', which count supersymmetric junctions between half-BPS line defects and a Neumann boundary condition in 4D ${\cal N} = 2$ supersymmetric Yang-Mills theory with gauge group $SU(2)$. Real Schur quantization \cite{Gaiotto:2024tpl,Gaiotto:2024osr} presents these half-indices as expectation values of $S^1$-wrapped line defects in an auxiliary unitary quantum-mechanical system. \\[2mm] 
As we will see, the duality dictionary identifies the Hamiltonian of the SYK model with the wrapped supersymmetric Wilson line in ${\cal N}=2$ SYM, while the matter chords between local operators in the SYK model map to wrapped 't Hooft lines or wrapped supersymmetric interfaces in the gauge theory setup.

While the double scaled SYK model and the auxiliary quantum system for 4d ${\cal N} = 2$ $SU(2)$ SYM with Neumann b.c. are evidently two very different physical systems, they share the property that (a special subclass of) their correlation functions are captured by means of a third soluble quantum mechanical system: \\[2mm]
{(C)} The partition function and correlators of holomorphic Wilson line operators in 3D $SL(2,\mathbb{C})$ Chern-Simons theory  on an interval $I$ times a disk $C_\bR$ with an irregular singularity at the origin of the disk and specific topological boundary conditions.
\\[2mm]
The equivalence between (B) and (C) follows from the results presented in \cite{Gaiotto:2024tpl,Gaiotto:2024osr}. The goal of this paper is to exhibit the equivalence of (B) = (C) with the double scaled SYK model (A) and, along the way, shed new light on the status of DSSYK as a candidate holographic dual of low-dimensional de Sitter gravity \cite{Susskind:2021esx,Susskind:2022bia,Narovlansky:2023lfz,Verlinde:2024znh,Verlinde:2024zrh,Rahman:2024vyg,Rahman:2024iiu}.

The triality correspondence between the three systems (A), (B) and (C) is built on a direct mathematical equality between parallel quantities on all sides. Besides the matching between the partition functions and correlation functions, we will find that the Hilbert space structure and operator algebra in all three cases is identical and can be organized in the form of an unitary representation of the skein algebra Sk$_q(SL(2),C)$, the algebra of CS Wilson line operators on $C$, where $C$ is the orientation double of $C_\bR$ and the algebra is equipped with a specific $*$-structure induced by the orientation-reversal map $\tau$. 

Recall that Sk$_q(SL(2),C)$ is the quantization of the Poisson algebra of holomorphic functions on the moduli space of flat $SL(2,\mathbb{C})$ connections on the curve $C$. This basis of functions is spanned by $SL(2,\mathbb{C})$ Wilson lines along one-cycles on $C$ \cite{Gaiotto:2024osr}, possibly ending on the irregular singularities. The $*$-structure identifies a real locus consisting of flat $SL(2,\mathbb{C})$ connections on $C_\bR$ which are unitary at the boundary. This structure is manifest in system~(C). 

Our triality identifies the curve $C$ with the class S curve for the 4d ${\cal N} = 2$ SYM theory and the skein algebra describes the operator algebra of circle-wrapped Wilson-'t Hooft loops \cite{Gaiotto:2024osr,Kapustin:2007wm,Alday:2009fs,Tachikawa:2015iba}.
On the SYK side, the skein algebra encapsulates the  chord rules that govern the DSSYK correlation functions \cite{Verlinde:2024znh}. The curve $C_\bR$ then naturally takes on the role of the spatial geometry of the candidate holographic dual to the DSSYK model. It is tempting to identify the disk $C_\bR$ with the static patch of a dual 3D de Sitter space with an observer located at the center of the static patch, c.f. \cite{Witten:2023xze,Narovlansky:2023lfz,Verlinde:2024znh}.

Before stating our results and presenting the concrete evidence in support of this new correspondence, we now first briefly introduce the three relevant quantum systems.

\subsubsection*{Double scaled SYK}
\vspace{-1mm}

The SYK model is a 1D quantum system built from $N$ Majorana oscillators with anti-commutator relations $\{\psi_i,\psi_j\} = 2\delta_{ij}$ interacting via a $p$-the order Hamiltonian \cite{kitaevTalks,Sachdev:1992fk,Maldacena:2016hyu}
\be
\label{hsyk}
H  =  {\imath}^{p/2}\! \sum^{}_{i_1\ldots i_p}\! J_{i_1\ldots i_p} \psi_{i_1} \ldots \psi_{i_p}
\ee
with gaussian random couplings $J_{i_1\ldots i_p}$ with variance
$
\langle  (J_{i_1 \ldots i_p})^2\rangle   =  {\J^2 }{{N \choose p}}^{-1}
$. 
We choose units such that the dimensionful SYK coupling $\J$ is set equal to 1.
We are interested in the correlators of the Hamiltonian and local operators in the {\it double scaling limit}
\be
{N\to \infty}, \quad  {p \to \infty} \qquad \lambda \equiv p^2/N = {\rm fixed}
\ee
The correlation functions that remain finite in this limit are those of local operators of the form ${\cal O}_\Delta = \sum_{i_1,..,i_{p'}} K_{i_1...i_p'} \psi_{i_1} ...\psi_{i_p'}$ with $\Delta \equiv p'/p$ and $K_{i_1...i_p'}$ another set of gaussian random numbers. We will call $\Delta$ the scale dimension of ${\cal O}_\Delta$. The correlators of the double scaled SYK (DSSYK) model depend only the scale dimensions and the dimensionless parameter \cite{Berkooz:2018jqr}
\be
\mfq = e^{-2\lambda}.
\ee
We will also employ the square root $q \equiv e^{-\lambda}$ to lighten some formulae.\footnote{We follow the DSSYK literature conventions, which are unfortunately opposite to the conventions~in~\cite{Gaiotto:2024tpl}}

The simplest class of correlators are the moments $\la H^n \ra = {\rm Tr}_{\cal H}(H^n)$ of the Hamiltonian.  
In the double scaling limit, these correlators have been explicitly computed in \cite{Berkooz:2018qkz,Berkooz:2018jqr}. They are given by the following integral expression
\bea \label{eq:sykEn}
 \qquad \ \la H^n \ra 
=
\int_0^\pi \! \frac{d\theta}\pi \, E(\theta)^n\, (\mfq;\mfq)_\infty (e^{2i\theta};\mfq)_\infty (e^{-2i\theta};\mfq)_\infty ,
\eea
with $E(\theta) = \frac{-2\cos\theta}{\sqrt{\lambda(1\!-\!\mfq)}}$.
Here  the integral over $\theta$ runs from $0$ to $\pi$ and  $(z; \mfq)_\infty \equiv \prod_{k=1}^\infty (1 - z \mfq^k)$. In particular,  the DSSYK partition function 
evaluated at infinite temperature reads 
\be
\label{zsyk}
Z_{\rm DSSYK}(\mfq)= {\rm Tr}_{\cal H} \,\mathbb{1} =    \int_0^\pi \! \frac{d\theta}\pi\, \, 
(\mfq;\mfq)_\infty (e^{2i\theta};\mfq)_\infty (e^{-2i\theta};\mfq)_\infty .
\ee  
The derivation of the above results involves evaluating the sum over the chord diagrams produced by performing all Wick contractions between the Hamiltonian insertions. This sum is governed by the simple combinatoric rule that associates a factor of $\mfq$ to every chord intersection \cite{Berkooz:2018jqr}. 

For later reference, we also quote the result for the infinite temperature two-point function of local operators of scaling dimension $\Delta$ \cite{Berkooz:2018jqr,Berkooz:2018qkz}\footnote{Here we use the abbreviated notation
$f(e^{\pm ix}) = f(e^{ix})f(e^{-ix})$, etc.}
\bea
\label{ztwo}
\bigl\la {\cal O}_\Delta{\cal O}_\Delta \bigr\ra & = & \int_0^\pi \frac{d\theta_1}\pi\frac{d\theta_2}\pi 
\frac{(\mfq^{2\Delta}; \mfq)_\infty (\mfq,e^{\pm 2i \theta_1} ; \mfq)_\infty (\mfq, e^{\pm 2i \theta_2} ; \mfq)_\infty}{(\mfq^\Delta e^{\pm i \theta_1\pm i \theta_2} ; \mfq)_\infty}.
\eea 
The formal operators with $\Delta =-1$ will be particularly important below. 
These do not correspond to actual physical operators in SYK but are formally defined as local operators with matter chords that associate a factor of $\mfq^{-1}$ to every chord intersection with a Hamiltonian chord and a factor of $\mfq^{1/2}$ for any intersection of with another $\Delta=-1$ matter chord.
When inserted in averaged correlation functions together with $H$ they generate a closed operator algebra that we will demonstrate takes the form of a familiar skein algebra.

In the above formulas, the expectation values are defined as a suitably normalized and averaged trace over the Hilbert space of the microscopic SYK model. Alternatively, we can view the averaged correlation functions as input data and then extract a Hilbert space via the GNS construction. 
In the following, we will refer to the latter notion as the Hilbert space of the DSSYK model. This Hilbert space and correlators are concisely captured with the help pf a $\mfq$-deformed oscillator algebra $[\mfa^\dag,\mfa]_\mfq=1$ and associated number operator $[\mfn,\mfa^\dag] = - [\mfn,\mfa]=1$, via the identifications (here ${\bf H} = \lambda^{1/2} H$)
\begin{align}
& { \bf H}= \mfa + \mfa^\dag, \qquad\qquad {\cal O}_{\Delta} {\cal O}_{\Delta} = \mfq^{\mfns\Delta} .
 \end{align}

In the following we will mostly restrict to the subsector $\cH_{\mathrm{DSSYK}}$ of the DSSYK Hilbert space spanned by acting with the Hamiltonian and an even number of operators ${\cal O}_{-1}$. This subsector is spanned by states obtained by acting with raising operators $\mfa^\dag$ on the vacuum of the $\mfq$-deformed harmonic oscillator. The $\mfq$-oscillators $\mfa$, $\mfa^\dag$ and the number operator $\mfq^{-\mfn}$ act within this subsector. 

The action of $\bf H$ and ${\cal O}_{-1}$ on this subsector defines an operator algebra $\cA_{\mathrm{DSSYK}}$ acting on a module ${\cal M}_{\mathrm{DSSYK}}$ consisting of finite linear combinations of $(\mfa^\dagger)^n|0\rangle$. The unitary structure of the problem is captured by the $*$-structure 
\begin{equation}
    \mfa^\dagger = \tau(\mfa) 
\end{equation}
on $\cA_{\mathrm{DSSYK}}$ and the compatible inner product on the module 
${\cal M}_{\mathrm{DSSYK}}$, so that the DSSYK Hilbert space $\cH_{\mathrm{DSSYK}}$ is the $L^2$ closure of ${\cal M}_{\mathrm{DSSYK}}$. 

\medskip

\subsubsection*{Schur correlation functions}
\vspace{-1mm}

Schur correlation functions are a special class of protected quantities in 4D  ${\cal N}=2$ supersymmetric gauge theory \cite{Gadde:2011ik,Gadde:2011uv,Dimofte:2011py,Drukker:2015spa,Cordova:2015nma,Cordova:2016uwk,Pan:2021mrw}. They can be defined as Witten indices that count supersymmetric local operators in the presence of Wilson-'t Hooft lines, or equivalently as supersymmetric partition functions on $S^3 \times S^1$ with line defects wrapping the $S^1$ \cite{Gang:2012yr}.\footnote{The former definition does not require the theory to be super-conformal. A sharper definition of supersymmetric local operators involves the {\it holomorphic-topological} twist \cite{Kapustin:2006pk,Mikhaylov:2017ngi}, see \cite{Gaiotto:2024tpl} for a discussion. For SCFTs, the Schur half index can be defined as a Witten index on the half-three sphere $HS^3\times S^1$ by the standard state-operator map. Pure $SU(2)$ gauge theory, however, is not superconformal. For non-conformal theories, one can use a state-operator map in the HT-twisted theory. This should correspond to some complicated rigid supergravity background for the physical theory on $HS^3\times S^1$.}  Of special relevance to our story if the {Schur half-index} \cite{Cordova:2016uwk}, formally 
defined as
\be
I\!\!I (\mfq) ={\rm Tr}_{\Op^\partial\strut} (-1)^{2 R} \mfq^{j_3 + R}
\ee
where $\Op^\partial$ denotes the linear space spanned by quarter-BPS  local operators supported at a boundary with supersymmetric Neumann boundary conditions. The Schur half-index counts all such local boundary operators, weighted with signs and a twisted spin fugacity specified by the sum of angular momentum $j_3$ along the symmetry axis of the half-three sphere and the $R$-charge $R$. The twist parameter $\mfq$  is a real number between 0 and 1 and will momentarily be identified with the SYK $\mfq$-parameter introduced above.

The first clue about the new quantum equivalence we propose is that the SYK partition function $Z_0(\mfq)$ given in \eqref{zsyk} is identical to the {Schur half-index} $I\!\!I_{\rm SW}$ of Seiberg-Witten theory \cite{Seiberg:1994aj,Seiberg:1994rs}, aka pure  4D  ${\cal N} = 2$ $SU(2)$ gauge theory, on a half-space $\bR^+\!\times \bR^3 $ with Neumann boundary conditions for the gauge fields:
\bea
\boxed{\;\, Z_{\rm DSSYK} (\mfq) = I\!\!I_{\rm SW} (\mfq)\small{}^{\strut}_{\strut}\;\, }
\eea
This match extends to a more complete set of identifications. More generally, the moments (\ref{eq:sykEn}) of the SYK Hamiltonian are identical (up to a trivial overall normalization) to a generalization of the Schur half-index of SW theory which counts all quarter-BPS local operators placed at a location where $n$ supersymmetric Wilson lines end together at the Neumann boundary. Furthermore, we will show that SYK correlators involving an even number of ${\cal O}_{-1}$ operator insertions  in the SYK model can also be reproduced on the gauge theory side by adding supersymmetric 't Hooft lines to the mix. 

It is useful to introduce a square-root $\fq = \mfq^{\frac12}$. 
The dependence of Schur correlation functions on the choice of supersymmetric line operators factors through the ``K-theoretic Coulomb branch algebra'' $\cA_{\mathrm{Schur}}$ of the 4d theory. This is an algebra over $\bZ[\fq,\fq^{-1}]$ which encodes the fusion of parallel line defects. 

In order to present the algebra, it turns out to be useful to enlarge $\cA_{\mathrm{Schur}}$ to an algebra $\cA'_{\mathrm{Schur}}$ that includes line defects with odd magnetic charge, which would carry a visible (but topological) central Dirac string in an $SU(2)$ gauge theory. The structure constants of $\cA'_{\mathrm{Schur}}$ involve square roots of $q$, which cancel out in $\cA_{\mathrm{Schur}}$. As algebra generators for $\cA'_{\mathrm{Schur}}$, we can choose the Wilson line $w_1$ in the spin 1/2 representation of $SU(2)$ and the elementary 't Hooft loop operator $\tH_0$ of charge $1$. More generally, we can have Wilson lines $w_n$ in the spin $n/2$ representation and dyonic loops $T_a$ with electric charge $a$ and unit magnetic charge. We will momentarily introduce line defects with higher magnetic charge. 

The Wilson lines form a commutative algebra given by the classical representation ring of $SU(2)$. The mixed operator product between Wilson and dyonic lines is non-commutative:\footnote{The physical origin of the non-commutativity is that the presence of the line operators generates a Poynting vector carrying the angular momentum $j_3$ proportional to the exterior product of the electro-magnetic charge vectors. The magnitude of this effect is independent of the distance between two lines but the sign depends on the ordering \cite{Gaiotto:2010be,Ito:2011ea,Tachikawa:2015iba}. 
The presence of the Dirac strings motivates the fractional powers of $\fq$. As soon as we restrict to $\cA_{\mathrm{Schur}}$, only integral powers will occur.}
\begin{align}
\label{wthalgebra}
  & w_1 w_n = w_{n+1} + w_{n-1}\\[.5mm]
 & w_1 \tH_a = \fq^{-\frac12} \tH_{a+1} + \fq^{\frac12} \tH_{a-1} \\[.5mm]
   & \tH_a w_1 = \fq^{\frac12} \tH_{a+1} + \fq^{-\frac12} \tH_{a-1} \\[.5mm]
   & \tH_a \tH_{a+1} = \fq \tH_{a+1} \tH_a   \\[.5mm]
   & \tH_{a-1} \tH_{a+1} = -1 + \fq \tH_{a}^2 \\[.5mm]
   & \tH_{a+1} \tH_{a-1} = -1 + \fq^{-1} \tH_a^2 
\label{wthalgebrat}
\end{align}
The line defects with higher magnetic charges are defined as 
\begin{equation}
    D_{m, m a + e} = \fq^{\frac12 e(e-m)}\tH_a^{m-e} \tH_{a+1}^e
\end{equation}
and complete the linear basis of $\cA'_{\mathrm{Schur}}$. The algebra 
$\cA_{\mathrm{Schur}}$ is generated by operators with even magnetic charge $m$.

The half-indices can also be decorated by boundary line defects, namely boundary Wilson lines $w^\partial_n$. These can be organized into a module ${\cal M}_{\mathrm{Schur}}$ for $\cA'_{\mathrm{Schur}}$ encoding the fusion of boundary and bulk lines. 
The Schur half-indices $I\!\!I_{m,m'}$ which count junctions between boundary lines define an inner product on ${\cal M}_{\mathrm{Schur}}$ which satisfies 
\begin{equation}
    I\!\!I_{m,a m'} = I\!\!I_{m,a,m'} = I\!\!I_{\tau(a)m,m'}
\end{equation}
where $a$ is a bulk like, $I\!\!I_{m,a,m'}$ counts junctions between $m$, $a$ and $m'$ and $\tau$ is a time-reversal symmetry of the system. 

Real Schur quantization is based on the claim that $I\!\!I_{m,m'}$ is a positive-definite inner product and thus the module ${\cal M}_{\mathrm{Schur}}$ can be promoted to a Hilbert space $\cH_{\mathrm{Schur}}$ \cite{Gaiotto:2024osr,Gaiotto:2024tpl}. 
Hermitean conjugation acts by 
\begin{equation}
    \tau(w_n) = w_n \qquad \qquad \tau(\tH_a) = \tH_{-a}
\end{equation}
Building on this, we will demonstrate that the operator algebras $\cA_{\mathrm{DSSYK}} \equiv \cA_{\mathrm{Schur}}$ match as $*$-algebras 
via $w_1 = \sqrt{1-\mfq}\, {\bf H}$ and ${\cal O}_{-1}{\cal O}_{-1}  = \tH_0^2$. Furthermore, ${\cal M}_{\mathrm{DSSYK}} \equiv \cM_{\mathrm{Schur}}$ as unitary representations of this algebra and thus $\cH_{\mathrm{DSSYK}} \equiv \cH_{\mathrm{Schur}}$. This is the first leg of the proposed triality. We will review details in the next sections.

\smallskip

\subsubsection*{$SL(2,\mathbb{C})$ Chern-Simons theory}
\vspace{-.5mm}

The algebra \eqref{wthalgebra}-\eqref{wthalgebrat} also has a sharp geometric meaning: it is the Skein algebra $\cA_{\mathrm{CS}}$ of space-like Wilson line operators in an $SL(2,\mathbb{C})$ Chern-Simons theory defined on the interval times the cylinder $C = \bC^*$ with rank $\frac12$ irregular singularities at $0$ and at $\infty$. Via the class S construction, this curve $C$ is associated to ${\cal N}=2$ $SU(2)$ SYM theory \cite{Witten:1997sc,Gaiotto:2009we,Gaiotto:2009hg}.

The quantization of $SL(2,\bC)$ Chern-Simons theory is an old and important problem \cite{Witten:1989ip,Dimofte:2010wxa}.  The Schur quantization formalism of  \cite{Gaiotto:2024osr} provides a powerful new approach that utilizes available knowledge about supersymmetric indices in ${\cal N}=2$ gauge theory to inform the computation of corresponding quantities in the CS theory. The theory with imaginary level is of particular interest and has a tantalizing connection to 3d Lorentzian de Sitter gravity \cite{Witten:1989ip}. This connection is one of our prime motivations for studying this class of quantum systems.

The phase space of $SL(2,\bC)$ Chern-Simons theory on the curve $C$ is the moduli space $\cM(SL(2),C)$ of complex flat connections on $C$, with appropriate behavior at the irregular singularities. 
Classically an irregular singularity is a localized defect on $C$ where the monodromy is decomposed into the product of a sequence of triangular Stokes matrices and a formal monodromy matrix. An irregular singularity of minimal rank $\frac12$ involves a single Stokes matrix, so that the monodromy is decomposed locally as 
\begin{equation}
    M = \begin{pmatrix} 0 & -1 \cr 1 & 0\end{pmatrix}\begin{pmatrix} 1 & \Tr \, M \cr 0 & 1\end{pmatrix} = \begin{pmatrix} 0 & -1 \cr 1 & \Tr \, M \end{pmatrix}
\end{equation}
In particular, the decomposition selects a specific state vector $|s\ra = {\,1\, \choose 0}$ in the spin 1/2 representation~of~$SL(2)$, together with an infinite sequence of images $M^n |s\ra$ such that consecutive pairs are linearly independent. 

In our case, we have two irregular singularities. So besides closed Wilson lines, we can define open path-ordered exponentials that end on the special vectors without violating gauge invariance:
\begin{align}
& w_n = {\rm{tr}}_{R_n}\Bigl( \rm{P} \exp \oint_{\gamma_A}\!\!\!A\Bigr),\qquad \qquad T_a = \la s'
| {\rm P} \exp\! \int_{\gamma_a}\!\!\!A |s\ra.
\end{align}
Here $\gamma_A$ denotes the A-cycle that separates the two irregular singularities and $\gamma_a$ is the open one-cycle that connects the two singularities while winding $a$ times around the A-cycle. In particular, the elementary operator $\tH_0$ corresponds to the open line operator with winding number zero. 

For later use, note that $|s\rangle$ can be promoted by symmetrization to a choice of vector in any finite-dimensional irrep~of~$SL(2)$ and thus gives a canonical endpoint for Wilson lines in the corresponding representations. With a bit more work, it can give natural endpoints for infinite-dimensional representations as well.

These classical definitions carry over to the quantum CS theory obtained by quantizing the Poisson algebra of classical Wilson lines. The presence of irregular singularities can be included in the quantization. The operator algebra relations \eqref{wthalgebra}-\eqref{wthalgebrat} can then be derived from the local skein rule \cite{kauffman1988new}
\\[-3mm]
\be
\label{skeinrule}
\qquad \raisebox{-3mm} {{\begin{tikzpicture}[rotate=45]
 \begin{scope}[yscale=.76,xscale=.76]
 \tikzset{
    partial ellipse/.style args={#1:#2:#3}{
        insert path={+ (#1:#3) arc (#1:#2:#3)}
    }
}      
        \draw[thick][black] (0, 0) -- (2, 0);       
        \draw[thick][black] (1, -1) -- (1, -.09);
        \draw[thick][black] (1, .09) -- (1, 1);
        \end{scope}
\end{tikzpicture}}\raisebox{5mm}{$\ \ \ = \ \ q^{\frac 1 2}$}~
 \raisebox{-5mm}{\begin{tikzpicture}[rotate=45]
\begin{scope}[yscale=.66,xscale=.66]
 \tikzset{
    partial ellipse/.style args={#1:#2:#3}{
        insert path={+ (#1:#3) arc (#1:#2:#3)}
    }
}         
       \draw[thick,black] (.5,-2.8) [partial ellipse=0:90:12mm and 12mm]; 
        \draw[thick,black] (2.8,-.5) [partial ellipse=180:270:12mm and 12mm];
        \end{scope}
\end{tikzpicture}}~~\raisebox{5mm}{$+ \ \ q^{-\frac 1 2}$\!\!\!\!\!}\raisebox{0mm}{
 \begin{tikzpicture}[rotate=45]
\begin{scope}[yscale=-.62,xscale=.66]
 \tikzset{
    partial ellipse/.style args={#1:#2:#3}{
        insert path={+ (#1:#3) arc (#1:#2:#3)}
    }
}      
    \draw[thick,black] (-.8,-2.6) [partial ellipse=0:90:12mm and 12mm]; 
        \draw[thick,black] (1.58,-.1) [partial ellipse=180:270:12mm and 12mm];
        \end{scope}
\end{tikzpicture}}}
\ee
which replaces the product of two crossing lines in CS theory with a linear sum of two non-crossing lines, together with some extra rules about lines ending on the same irregular singularity. Here $q$ is a square root of
\bea
\mfq = e^{\frac{4\pi i} \kappa} .
\eea  
in terms of the (critically shifted) level of the CS theory. \footnote{There are some subtleties we glossed over. The sections $s$ and $s'$ are normalized up to a sign flip, so only products of an even number of $T_a$'s are well-defined in $\cA_{\mathrm{CS}}$. The $T_a$'s define a larger algebra $\cA'_{\mathrm{CS}}$. The algebra $\cA_{\mathrm{CS}}$ only involves integral powers of $q$.}

We were intentionally ambiguous about the choice of real form of the gauge group. The choice must be compatible with the hermiticity properties of the generators of $\cA_{\mathrm{CS}}$. We claim that, to obtain a match with the decorated Schur half-indices in SW theory and with DSSYK, the correct choice is to use a reflection trick: we consider $SL(2,\bC)$
CS theory on a disk $C_\bR = C/\tau$ (see figure 4) with a special topological boundary condition where the gauge group is reduced to $SU(2)$. This setup is well-defined for imaginary level $k=i\kappa$, i.e. real $q$. 

The phase space on the $\bZ_2$ quotient $C_\bR = C/\tau$ is a real locus 
$\cM(SL(2),C_\bR)$ within the moduli space $\cM(SL(2),C)$ of flat $SL(2,\mathbb{C})$ connections on $C$. This reality restriction is the classical implementation of the reality constraint $a^\dagger = \tau(a)$.
On $C_\bR$, the Wilson line $w_n$ measures the holonomy along a closed loop  along the boundary of $C_\bR$ encircling the irregular singularity. The 't Hooft loop $\tH_0$ transports the vector $|s\rangle$ from the origin to the boundary and takes the $SU(2)$-invariant norm available there. The more general $\tH_a$ takes the $SU(2)$-invariant inner product of $|s\rangle$ and its image after winding $a$ times around the origin. 

In section 3 we will specify an auxiliary three-manifold $M_3$ with boundary $C$ designed such that the $SL(2)$ skein module ${\cal M}_{\rm CS}$ on $M_3$ embeds naturally in the Hilbert space associated to the $SL(2,\mathbb{C})$ CS theory on $C_\bR$ and matches with the module ${\cal M}_{\rm Schur} ={\cal M}_{\rm DSSYK}$ obtained from the other two corners of the triality. From this CS perspective, the quantization step amounts to building a Hilbert space $\cH_{\mathrm{CS}}$ algebraically from a collection of boundary states equipped with a natural inner product compatible with the action and hermiticity conditions of the Skein algebra ${\rm Sk}_q(C,\mathfrak{sl}_2)$. 

This provides the third leg of our triality.  Altogether, our results relate the DSSYK model to Schur correlators in SW theory and to $SL(2,\bC)$ Chern Simons theory with a novel unitary structure, and perhaps to 3d de Sitter gravity.

\medskip
\smallskip

This paper is organized as follows. In section 2  we review the definition of the Schur half-index and introduce the notion of real Schur quantization and correlators, specializing to 4D $SU(2)$ Seiberg-Witten theory. In section 3, we present the dual 6D and 3D perspective on the Schur correlators and their connection with the Skein algebra $\mathrm{Sk}_q(C,\mathfrak{sl}_2)$ of line operators in $SL(2,\mathbb{C})$ Chern-Simons theory on $C_\bR$. In section 4, we use the explicit match between Schur and DSSYK correlators to show that the Hilbert space of the double scaled SYK model forms a representation of this Skein algebra $\mathrm{Sk}_q(C,\mathfrak{sl}_2)$. We comment on how one may construct Schur correlators that match SYK correlators with matter.  We end with some open questions and future directions.

\bigskip

\section{Schur correlators in $SU(2)$ Seiberg-Witten theory}

\vspace{-1mm}
In this section we review the general features of Schur correlation functions and specialize to 4d $SU(2)$ Seiberg-Witten theory \cite{Seiberg:1994aj,Seiberg:1994rs} and Neumann boundary conditions. 

As an asymptotically free theory with both a UV lagrangian and a class ${\cal S}$ description, pure ${\cal N} = 2$ $SU(2)$ gauge theory is well understood both in the UV and in the IR. However, since it is not a superconformal field theory, the definition of the supersymmetric Schur index and correlation functions requires some special care. We will for the most part try to steer clear of these subtleties and refer to \cite{Gaiotto:2024osr} for a more detail discussion. Our main focus will be on those elements that are instrumental in the match between ${\cal N} = 2$ SYM theory and the double scaled SYK model. 

\subsection{Schur index}
\vspace{-1mm}

The Schur index ${\cal I}(q)$ of a superconformal ${\cal N}=2$ gauge theory can be defined abstractly by a trace over the Hilbert space of states on the three sphere or, by the state operator correspondence, over the space of (supersymmetric) local operators. The trace is twisted by $(-1)^{2R_3} \fq^{2j_3+2R_3}$ with $j_3$ the rotation generator on the $S^3$ and $R_3$ one of the $SU(2)$ R-charges \cite{Kinney:2005ej,Gadde:2011ik,Cordova:2015nma}. Here $\fq= \mfq^{\frac12}$ is a formal parameter that we will choose to be a real number such that $0<\fq^2<1$.

This index ${\cal I}(q)$ is a protected quantity and does not depend on the geometry or the gauge coupling. For Lagrangian field theories, the Schur index can be computed by taking the coupling constant to zero and can be understood as counting gauge-invariant local operators built from BPS letters \cite{Gadde:2011ik}. The result can be then expressed by a simple matrix integral \cite{Gadde:2011uv}.

For a gauge theory with one vector multiplet and $n_F$ $\frac 1 2$-hypermultiplets transforming some representation $R$, the Schur index  takes the form
\be
{\cal I}(q) = \int dU \, {\rm PE}\bigl[f^V(q) \chi_{\rm adj} (U) + n_F f^{\frac 1 2 H}(q)\chi_R(U)\bigl]
\ee 
with $f^V(q)$ and $f^{\frac 1 2 H}(q)$ the single letter partition functions for the vector and $\frac 1 2$-hypermultiplet. 
\be
f^V(q) = \frac{-2q^2}{1-q^2}\qquad \qquad f^{\frac 12 H} = \frac{q}{1-q^2},
\ee
Here PE denotes the plethystic exponential
\be
{\rm PE}\bigl[f(q,z)\bigr] \equiv \exp\biggl[\,\sum_{n=1}^\infty \frac{1}{n} f(q^n,z^n)\biggr]
\ee
 and $\chi_{\rm adj}(U)$ and $\chi_{R}(U)$ denote the characters in the corresponding representations. One can think of PE as 
implementing a second quantization step. 

Pure ${\cal N}=2$ gauge theory is not superconformal. The above definition of the Schur index as the twisted supersymmetric partition on $S^3 \times S^1$ can in principle be generalized to non-superconformal theories by placing the theory on some appropriate supergravity background, or by employing a holomorphic-topological twist \cite{Kapustin:2007wm,Mikhaylov:2017ngi}. We will not elaborate on this \cite{Gaiotto:2024tpl}, but borrow from the HT twist a splitting of flat Euclidean space-time into a product 
 \be
 M_4 = \bR^2 \times \bC.  
 \ee
 The $\bC$ factor is rotated by the $j_3$ generator appearing~above. The operators we count live at the origin.

Specializing to pure $SU(2)$ gauge theory, the above formulas for the Schur index simplifies to
\be
{\cal I}(q) = \int\!\frac{d\theta}{\pi}  \sin^2 \theta \, {\rm PE} \bigl[f^V(q) \chi_2(\theta)\bigr]
\ee
with $\chi_2(e^{i\theta}) = e^{2i\theta} + e^{-2i\theta} +1$. Using standard identities for the plethystic exponential, we can express the right-hand side in terms of an integral over Pochhammer symbols
\be
\label{schurone}
{\cal I}(q) = \int\!\frac{d\theta}{4\pi}  \sin^{2} \theta \, \bigl[(q^2;q^2)_\infty (q^2e^{\pm 2i\theta};q^2)_\infty \bigr]^2 .
\ee
where the integral runs from $0$ to $\pi$. Each factor in the infinite product accounts for an individual ``letter'' in the gauge-invariant operators: holomorphic derivatives of two types of gauginoes. The $\theta$ integral implements a projection onto $SU(2)$-invariant operators.

The integrand looks roughly like the square of the integrand in the partition function of the double scaled SYK model. In order to get a precise match, we will consider the Schur half-index, which counts protected boundary local operators for a half-BPS boundary condition, and decorate it by a collection of supersymmetric line defects ending at the origin.

\subsection{Schur half-index} 
\vspace{-1mm}

We now place the ${\cal N}=2$ gauge theory on a 4D space-time 
\be
M^+_4 = \mathbb{R}^+\! \times \mathbb{R} \times \mathbb{C}
\ee
with boundary $\partial M^+ = \mathbb{R} \times \mathbb{C}$ \cite{Dimofte:2011py, Cordova:2016uwk}. The gauge theory has a large variety of useful boundary conditions. We choose Neumann boundary conditions for the gauge fields. SUSY fixes the boundary conditions for all other fields \cite{Drukker:2010jp}. In particular, the boundary conditions remove roughly half of the letters from the index and results in the expression 
\be
\label{schurhalf}
I\!\!I(q) = \int\!\frac{d\theta}{4\pi}  \sin^{2} \theta \, (q^2;q^2)_\infty (q^2e^{\pm 2i\theta};q^2)_\infty  \,.
\ee
Again, the $\theta$ integral imposes a projection on gauge-invariant operators. This can be modified to a projection onto operators which transform in an irrep of $SU(2)$ of dimension $2n+1$: 
\be
\label{schuronet}
I\!\!I_{w_n}(q)  = \int\!\frac{d\theta}{4\pi}  \sin^{2}\! \theta \, \mu(\theta) \,\chi_n(\theta)
\ee
with 
\be
\label{mutheta}
\mu(\theta) = (q^2,q^2)_\infty (q^2 e^{\pm 2i\theta_1};q^2)_\infty =  \frac{\vartheta_1(2\theta,q^2)}{\sin \theta}
\ee
This modified index has a simple interpretation: it counts boundary local operators placed at the end of an half-BPS Wilson line defect for the same representation \cite{Dimofte:2011py}. The defect is placed along a ray in $\bR^+ \times \bR$. The specific slope is immaterial: we can consider a bulk Wilson line $w_n$ transverse to the boundary or merge it with the boundary on either side of the origin to give rise to a boundary Wilson line $w^\partial_n$.

\subsection{Half-indices decorated by line defects}

\vspace{-1mm}
A very general configuration which allows for a Schur-like count of protected local operators at the origin involves an half-BPS boundary condition, giving the $M^+_4$ geometry, a pair of boundary line defects $L^\partial_m$ and $L^\partial_{m'}$ placed along the two halves of the $\bR$ factor in the geometry and a sequence of half-BPR bulk line defects $L_{a_i}$ placed along a sequence of rays in $\bR^+ \times \bR$. 

The space $\mathrm{Op}_{m,a_1, \cdots a_k,m'}$ of protected local operators intertwining all of these defects depend on the choice of defects and on the relative order of their slopes. We can define the usual Witten index 
\be
I\!\!I_{m,a_1, \cdots a_k,m'}(q) = {\rm Tr}_{\mathrm{Op}_{m,a_1, \cdots a_k,m'}}(-1)^{2 R_3} q^{2j_3 + 2R}
\ee
Here $j_3$ denotes the rotation operator of $\bC$ and $R$ denotes a Cartan generator of $SU(2)_R$. 

\begin{figure}[t]
\centering
\raisebox{2.2cm}{\large $a)$}~~~~\begin{tikzpicture}[rotate=0,xscale=1.28,yscale=1.28] 
\tikzset{
    partial ellipse/.style args={#1:#2:#3}{
        insert path={+ (#1:#3) arc (#1:#2:#3)}
    }
}   
\draw[magenta,very thick,decorate, decoration={snake, segment length=1.5mm, amplitude=.1mm}] (0,0) -- (.35,.7);
\draw[magenta] (.44,.94) node {\scriptsize $L_1$};
\draw[magenta,very thick,decorate, decoration={snake, segment length=1.5mm, amplitude=.1mm}] (.52,1.04) -- (.88,1.78);
\draw[cyan,very thick,decorate, decoration={snake, segment length=1.5mm, amplitude=.1mm}] (.52,-1.04) -- (.62,-1.24);
\draw[cyan,very thick,decorate, decoration={snake, segment length=1.5mm, amplitude=.1mm}] (0,0) -- (.35,-.7);
\draw[cyan] (.4,-.9) node {\scriptsize $L_{k-1}$};
\draw[\darkgreen,thick,decorate, decoration={snake, segment length=1.5mm, amplitude=.1mm}] (0,0) -- (-.35,.7);
\draw[\darkgreen] (-.4,.88) node {\scriptsize $L_2$};
\draw[\darkgreen,very thick,decorate, decoration={snake, segment length=1.5mm, amplitude=.1mm}] (-.52,1.04) -- (-.88,1.78);
\draw[blue,very thick,decorate, decoration={snake, segment length=1.5mm, amplitude=.1mm}] (0,0) -- (-.72,.24);
\draw[blue] (-.92,.32) node {\scriptsize $L_3$};
\draw[blue,very thick,decorate, decoration={snake, segment length=1.5mm, amplitude=.1mm}] (-1.08,.36) -- (-1.83,.61);
\draw[\darkred, very thick,decorate, decoration={snake, segment length=1.5mm, amplitude=.1mm}] (0,0) -- (.72,-.24);
\draw[\darkred] (.92,-.32) node {\scriptsize $L_k$};
\draw[\darkred, very thick,decorate, decoration={snake, segment length=1.5mm, amplitude=.1mm}] (1.08,-.36) -- (1.83,-.61);
\draw[blue,->] (1.1,0) arc (0:45 :1) node[midway, right]  {{\footnotesize $\varphi$}};
\draw[thick] (0,0) circle (.45mm);
\draw[thick,\darkred] (-.92,-.58) circle (.15mm);
\draw[thick,\darkred] (-.8,-.8) circle (.15mm);
\draw[thick,\darkred] (-.58,-.92) circle (.15mm);
\draw[thin,dotted] (0,0)--(2,0);
\draw (0,-1.70) node {${M_4} =\, \mathbb{R}^2 \times \mathbb{C} $};
\end{tikzpicture}~\raisebox{1.95cm}{}~~~~~~~~~~~~~~~~~\raisebox{.05cm}{\begin{tikzpicture}[xscale=1.26,yscale=1.26] 
\tikzset{
    partial ellipse/.style args={#1:#2:#3}{
        insert path={+ (#1:#3) arc (#1:#2:#3)}
    }
}   
\draw[magenta,very thick,decorate, decoration={snake, segment length=1.5mm, amplitude=.1mm}] (0,0) -- (.44,.66);
\draw[magenta] (.52,.78) node {\scriptsize $L_1$};
\draw[magenta,very thick,decorate, decoration={snake, segment length=1.5mm, amplitude=.1mm}] (.6,.9) -- (1.12,1.68);
\draw[\darkgreen,very thick,decorate, decoration={snake, segment length=1.5mm, amplitude=.1mm}] (0,0) -- (-.1,.8);
\draw[\darkgreen] (-.15,.95) node {\scriptsize $L_2$};
\draw[\darkgreen,very thick,decorate, decoration={snake, segment length=1.5mm, amplitude=.1mm}] (-.14,1.12) -- (-.24,1.92);
\draw[blue,very thick,decorate, decoration={snake, segment length=1.5mm, amplitude=.1mm}] (0,0) -- (-.7,.35);
\draw[blue] (-.88,.44) node {\scriptsize $L_k$};
\draw[blue,very thick,decorate, decoration={snake, segment length=1.5mm, amplitude=.1mm}] (-1.04,.52) -- (-1.78,.89);
\draw[blue,->] (1.1,0) arc (0:35 :1) node[midway, right]  {{\footnotesize $\varphi$}};
\draw[thick,\darkgreen] (-.91,.91) circle (.15mm);
\draw[thick,\darkgreen] (-.80,1.02) circle (.15mm);
\draw[thick,\darkgreen] (-.66,1.1) circle (.15mm);
\draw[very thick] (-2,0.01)--(2,0.01);
\draw[very thick] (-2,-0.01)--(2,-0.01);
\draw (0,-0.40) node {\small $\partial {M^+_4}$};
\draw (0,-1.60) node {${M^+_4} =\, \mathbb{R}^+\! \times \mathbb{R}  \times \mathbb{C} $};
\end{tikzpicture}
}
\caption{The geometric configuration of line defects that specify the Schur index (left) and half-index (right) decorated by an ordered set of line defects $L_{a_i}$.}
\end{figure}
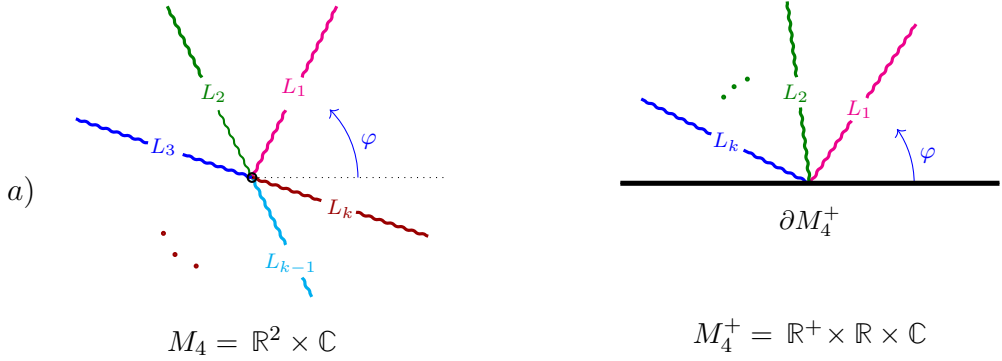

If two slopes are deformed to the same value, the corresponding defects are fused in a supersymmetric manner. The fusion of bulk lines gives a non-commutative operator algebra, the equivariant K-theoretic Coulomb branch algebra $\cA$ of the 4d theory. The fusion of a bulk and a boundary line in either direction defines a left- and a right- $\cA$-modules $\cM$ and $\bar \cM$ of boundary lines. A small but important subtlety is that the change in the slopes of the bulk lines is accompanied by a change in the SUSY algebra they preserve. This results in an anomalous $U(1)_r$ rotation which affect the charge labelling of line defects by Witten effect \cite{Gaiotto:2010be}. We will discuss this in greater detail when needed. 

As a result, $I\!\!I_{m\,a_1 \cdots a_k\,m'}(q)$ can be expressed in multiple ways in terms of half-indices decorated by boundary lines only:
\begin{equation}
    I\!\!I_{m,a_1, \cdots a_k,m'}(q) = I\!\!I_{(m a_1 \cdots a_i),(a_{i+1} a_k m')}(q)
\end{equation}
for all choices of $i$. Mathematically, $I\!\!I$ defines a linear function on the tensor product $\bar \cM \otimes_{\cA} \cM$.

Real Schur quantization pertains to half-indices for boundary conditions which are preserved by some time-reversal symmetry $\tau$ \cite{Gaiotto:2024osr}. The time-reversal symmetry give maps $\tau:\cM \to \bar \cM$ and $\tau:\cA \to \cA^{\mathrm{op}}$, with $\tau^2=1$, so that 
\begin{equation}
    \tau(a_1 a_2) = \tau(a_2) \tau(a_1) \qquad \qquad \tau(a m) = \tau(m) \tau(a)
\end{equation}
Conjecturally, 
\begin{equation}
    I\!\!I_{\tau(m) m}>0 \qquad \qquad -1<q<1
\end{equation}
and thus the half-indices define a positive-definite inner product on $\cM$:
\begin{equation}
    \langle m|m'\rangle \equiv I\!\!I_{\tau(m) m'}
\end{equation}
This satisfies 
\begin{equation}
    \langle a\,m|m'\rangle=I\!\!I_{\tau(a\,m),m'}=I\!\!I_{\tau(m),\tau(a)m'} = \langle m|\tau(a)\,m'\rangle 
\end{equation}
i.e. $\tau(a)^\dagger = a$. We can thus define an Hilbert space $\cH$ as a completion of $\cM$. It carries an unitary $\cA$ action. The half-indices are interpreted as expectation values: 
\begin{equation}
    I\!\!I_{\tau(m)\,a_1 \cdots a_k\,m'}(q) = \langle m|a_1 \cdots a_k|m'\rangle\strut \,
\end{equation}

We can now specialize to Neumann boundary conditions for Seiberg-Witten theory. The algebra $\cA$ is linearly generated by (K-theory classes of) 't Hooft-Wilson lines $D_{m,e}$ with magnetic charge $m$ and electric charge $e$, with $D_{m,e} = D_{-m,-e}$. In particular, we denote $D_{0,e} = w_e$ and $D_{1,e} = T_e$. The module $\cM$ is linearly generated by boundary Wilson lines $w^\partial_n$. The time-reversal 
symmetry acts as $\tau(D_{m,e}) = \tau(D_{m,-e})$.

The computation of the algebra and module action requires a careful account of monopole bubbling effects. There is an useful description of boundary lines as functions $\chi_n(\theta)$ of $\theta$ and of $D_{m,e}$ as certain difference operators in $\theta$. Then the Schur half-indices can be computed by acting with all the difference operators
on the boundary lines and then using  
\be
\label{schuronep}
I\!\!I_{w^\partial_n,w^\partial_{n'}}(q)  = \int\!\frac{d\theta}{\pi}  \sin^{2}\! \theta\, \mu(\theta) \,\chi_n(\theta)\chi_{n'}(\theta)
\ee
with $\mu(\theta)$ given in \eqref{mutheta}. 

The half-index 
counting boundary local operators in the presence of $n$ Wilson lines in the spin 1/2 representation
\be
\label{nwilson}
    I\!\!I_{w_1^n}(q) =  I\!\!I_{1^\partial,{w_1,\cdots,w_1},1^\partial}(q) \strut
\ee
is given by the same expression, up to a trivial overall normalization, as the $n$-th moment of the Hamiltonian in the double scaled SYK model. This is the result announced in the introduction.

For completeness, we briefly review {\it complex} Schur quantization. This concerns indices ${\cal I}_{a_1, \cdots a_k}$ counting local junctions of a sequence of line defects in $\bR^2 \times \bC$. These indices satisfy
\begin{equation}
    {\cal I}_{a_1, \cdots a_k} = {\cal I}_{\rho^2(a_k), \cdots a_{k-1}} 
\end{equation}
for a certain automorphism $\rho$ of $\cA$ which encodes $\pi$ slope rotations. It thus gives a {\it twisted trace} on the algebra $\cA$. 
The positivity condition 
\begin{equation}
    {\cal I}_{\rho(a) a} > 0 \qquad \qquad 0<q^2<1
\end{equation}
allows one to define an Hilbert space $\cH_\bC$ as the closure of $\cA$. It has an unitary action of a double $\cA \otimes\cA^{\mathrm{op}}$ with a $*$ structure exchanging the two factors. 

The automorphism $\rho$ will play a small role in our analysis as well. Indeed, it allows one to better compare the left- and right- actions of $\cA$ on boundary lines. The 
combination $\tau\circ \rho$ will preserve the $\cA$ action on ${\cal M}$
up to the replacement $q \to q^{-1}$. 

We identify $\tau$ as a time-reversal symmetry of the Seiberg-Witten theory with Neumann boundary conditions. Then the combination $\tau\circ \rho$ can be understood as rotating a line defect which is almost fusing to the boundary along the future ray to a line defect which is almost fusing to the boundary along the past ray and then reflecting it back to the future. This should be a symmetry of the fusion process, as observed. 

The half-index can also be interpreted as a distributional 
state in $\cH_\bC$, simply because it is a linear function on $\cA$. We will not use this property, as our main focus is the full Hilbert space $\cH$. It can be an useful computational tool in some situations, though. 

In this paper, instead, we will use other half-BPS boundary conditions or interfaces which admit a BPS junction with Neumann b.c. as a tool to produce interesting distributional states or kernels in $\cH$. We will come back to this in a later Section as we seek a gauge-theoretic interpretation of local scaling operators ${\cal O}_\Delta$ in DSSYK.  

\subsection{Explicit construction of $\cA$ and $\cM$}
The difference operator description of $D_{a,b}$ is naturally expressed in terms of auxiliary operators $u$ and $v$, which satisfy the Weyl algebra
\begin{equation}
    u v = q \,v u
\end{equation}
We can identify $v = e^{i \theta}$ and $u$ as a shift of $\theta$ in the imaginary direction when acting on the characters $\chi_n(\theta)$ representing boundary Wilson lines. 

We write 
\begin{equation}
    w_n = v^n + v^{n-2} + \cdots + v^{-n}
\end{equation}
so that a bulk Wilson line brought to the boundary becomes a boundary Wilson line. The dyonic loop action is richer: 
\begin{equation}
    \tH_a = \frac{q^{\frac{a-1}{2}}}{v^{-1}-v} (v^{a-1} u - v^{1-a} u^{-1})
\end{equation}

With some patience, we can reconstruct the relations (\ref{wthalgebra})-(\ref{wthalgebrat}). 
We can also define operators with higher magnetic charge as
\begin{equation}
    D_{k_1+k_2,(k_1+ k_2)a + k_2} \equiv q^{-\frac{k_1 k_2}{2}}\tH_a^{k_1} \tH_{a+1}^{k_2} 
\end{equation}
The relations \eqref{wthalgebra} are sufficient to simplify the product of any linear generators of $\cA'$. 

Recall that in the following we are only interested in the sub-algebra 
$\cA$ of operators with even magnetic charge. 

We now describe the action of bulk lines on boundary lines $w_n^\partial$ as a module $\cM$ for $\cA$. We can immediately compute two crucial module relations:
\begin{align}
    w_n \cdot 1^\partial &= w_n^\partial \cr
    \tH_1 \cdot 1^\partial &= 0 \\
    \tH_0 \cdot 1^\partial &= \fq^{-\frac12} 1^\partial\nonumber
\end{align}
More generally, 
\begin{align}
    \tH_{a+2}\cdot 1^\partial &= \frac{q^{\frac{a+1}{2}}}{v^{-1}-v} (v^{a+1} - v^{-a-1}) 1^\partial  = -q^{\frac{a+1}{2}} w_a^\partial \\
    \tH_{-a} \cdot 1^\partial &= \frac{q^{\frac{-a-1}{2}}}{v^{-1}-v} (v^{-a-1} - v^{1+a}) 1^\partial =q^{\frac{-a-1}{2}}w_a^\partial\nonumber
\end{align}
These relations fully determine the action of $\cA'$ and thus of $\cA$ on $\cM$. 

The inner product \eqref{schuronep} induces an Hermitean structure on the algebra. The measure is almost invariant under the $u$ shifts, up to a crucial overall monomial in $q$ and $v$ which insures 
\begin{equation}
    w_1^\dagger = w_1 \qquad \qquad \tH_a^\dagger = \tH_{-a}
\end{equation}
It turns out that $\rho(w_n) = w_n$ and $\rho(\tH_a) = -\tH_{a-2}$. Then $\tau(\rho(\tH_a)) = - \tH_{2-a}$
which indeed intertwines the $\cA'$ action on $\cM$ for $q$ and $q^{-1}$, as expected. 

For later use, notice that the inner product on $\cM$ is essentially determined by 
the algebraic structure and Hermiticity conditions:
\begin{equation}
    I\!\!I_{w_n}(q) = q^{\frac{n+1}{2}}I\!\!I_{\tH_{-n}}(q)= q^{\frac{n+1}{2}}I\!\!I_{\tH_{n}}(q)= - q^n I\!\!I_{w_{n-2}}(q)
\end{equation}
which together with $I\!\!I_{w_1}(q)=I\!\!I_{\tH_1}(q)=0$
leads to 
\begin{equation}
    I\!\!I_{w_{2n}}(q) = (-1)^n q^{n(n+1)}I\!\!I(q)
\end{equation}
and 
\begin{equation}
    I\!\!I_{w_{2n+1}}(q) =0
\end{equation}
This agrees with the integral expression for the half-indices. As the characters $\chi_n(\theta)$ are a basis of the Hilbert space, ${\cal M}$ is dense in $L^2([0,\pi], d\mu) = \cH$. 

We have thus executed successfully the real Schur quantization algorithm for Neumann boundary conditions in SW theory and produced an unitary representation of the $*$-algebra $\cA$ on $L^2([0,\pi], d\mu)$
and an useful module $\cM$ whose closure is the Hilbert space. 

\subsection{What are we quantizing?}
The $q \to \pm 1$ limits  of the algebra $\cA$ are commutative Poisson  algebras $\cA^\pm$ and can be interpreted as Poisson algebras of functions on 
complex symplectic manifolds $\cM^\pm$. The map $\tau$ defines a complex 
involution of $\cM^\pm$ which fixes a real locus $\cM^\pm_\bR$. We identify this locus with the real phase space which is quantized by real Schur quantization.

The constraints satisfied by $1^\partial$ also define a sub-manifold of 
$\cM^\pm$: a complex Lagrangian sub-manifold $\mathcal{L}^\pm$ which is fixed by 
the combination $\tau\circ \rho$. 
At first sight, it is curious that $\tau$ appears here with two roles: it both controls the choice of real locus $\cM^\pm_\bR$ in $\cM^\pm$ and the choice of complex Lagrangian manifold we use to produce a vacuum state. This appears to be a general feature of both real Schur quantization \cite{Gaiotto:2024osr} and real sphere quantization \cite{Gaiotto:2023hda}. 

More precisely, recall that $\cM^\pm$ is actually an hyper-K\"ahler manifold. The $U(1)_r$ anomalous rotation associated to $\rho$ can be understood as a hyper-K\"ahler rotation, which changes the choice of complex structure on the underlying manifold. We expect $\cM^\pm_\bR$ and $\mathcal{L}^\pm$ to be related by an hyper-K\"ahler rotation by $\pi/2$,
relating the ``$J$'' and ``$K$'' complex structures on the underlying manifold. 

We can attempt to understand this phenomenon by looking at Schur quantization as a form of ``brane quantization'' based on the A-model 
on $\cM^\pm$ \cite{Gukov:2008ve}. The brane quantization of $\cM^\pm_\bR$ involves a compactification of the A-model on a segment, with a $B_{\mathrm{cc}}$ boundary condition and a boundary condition supported on $\cM^\pm_\bR$. A state can be created as a boundary state for a third boundary condition supported on 
$\mathcal{L}^\pm$ \cite{Gaiotto:2021tsq}. This requires a choice of ``corners'' where the $\mathcal{L}^\pm$ brane encounters the $B_{\mathrm{cc}}$ and $\cM^\pm_\bR$ branes. 
The first choice of corner is an element of $M$. 

The second choice is what we need to understand here. We want to ``smoothly'' interpolate between $\cM^\pm_\bR$ and $\mathcal{L}^\pm$. The hyper-kahler rotation produces a family of A-branes which do so in a canonical way, demonstrating that $\cM^\pm_\bR$ and $\mathcal{L}^\pm$ are secretly the ``same'' brane. 
A torus reduction of the Schur half-index does appear to give such an A-model configuration. Hence we see real Schur quantization as a tool to compute the A-model amplitudes required to make brane quantization explicit \cite{Gaiotto:2024osr}.

\section{Class ${\cal S}$ description and $SL(2,\mathbb{C})$ Chern-Simons theory}

\vspace{-2mm}
In this Section we review the 6d origin of SW theory and the relation between real Schur quantization and complex Chern-Simons theory on a surface with boundary. 
We use the term $SL(2,\mathbb{C})$ Chern-Simons theory
to refer to a TQFT defined by a path integral over the space of $SL(2,\mathbb{C})$ connections $A$ with an action which is a linear combination of the real and imaginary parts of the Chern-Simons action of $A$. The level in front of the real part is quantized and we set it to $0$. The level in front of the imaginary part $\kappa$ is not quantized and we take it to be generic.

\subsection{Class ${\cal S}$ description}
\vspace{-1mm}

The class ${\cal S}$ description of 4D ${\cal N}=2$ supersymmetric $SU(2)$ gauge theory is obtained \cite{Gaiotto:2009we} by compactifying the 6D (2,0) superconformal field theory associated to the $\mathfrak{sl}_2$ Lie algebra on a Riemann surface $C$
given by a sphere with two irregular singularities \cite{Gaiotto:2009hg}, specified momentarily. 
The class ${\cal S}$ description extends to include the line defects which are our main interest. The 6d theory has some half-BPS surface defects, which can be wrapped on curves in $C$ to define 't Hooft-Wilson line defects in 4d. The curves may be closed or end on irregular singularities in specific ways.

It is worth pointing out that the class ${\cal S}$ description is compatible with the holomorphic-topological twist we used explicitly or implicitly in the 4d analysis: we expect the 6d theory to admit a HT twist on $\bR^4 \times \bC$ which reduces topologically along $C$ to give the 4d twisted gauge theory. The surface defects should become topological in such a twist. The same 6d twist is relevant, say, for the physical definition of knot topology in terms of surface defects \cite{Witten:2011zz}. Both in our context and in the knot theory setting, one writes $\bR^4 = \bR^2 \times \bC$ and typically places all defects at the origin of the complex plane, working equivariantly under rotations that keep the origin fixed. The surface defects are placed at the fixed point locus and extended along one topological direction in 4D and wrap one-cycle on $C$. The twisted rotation fugacity ``$\fq$'' which appears in the Schur index is literally the same as the ``$\fq$'' which appears in skein relations in (analytically continued) $SL(2,\bC)$ Chern-Simons theory.

A simple duality chain demonstrates that the same rotation-equivariant compactification which maps line defects to elements of $A_{\mathrm{Schur}}$ will map the surface defects to Wilson lines in Chern-Simons theory and thus identify $A_{\mathrm{Schur}}$ as the {\it Skein algebra} 
\begin{equation}
    A_{\mathrm{CS}} \equiv \mathrm{Sk}_q(C,\mathfrak{sl}_2)
\end{equation}
of space-like Wilson line defects in a CS theory. 
The classical version of the Skein algebra is the Poisson algebra of algebraic functions on the character variety of flat $SL(2)$ connection on $C$. Conversely, the Skein algebra is a canonical quantization of that Poisson algebra and can be recovered in many different ways. 

In the particular situation at hand, the space of flat connections on $C$ has a nice ``Fenchel-Nielsen'' coordinate system with a well-known quantization, which simply coincides with the auxiliary $u$, $v$ operator algebra we used in the previous Section. 
The auxiliary expressions for $w_1$ and $\tH_a$ precisely match the standard quantization of the corresponding generators of the Skein algebra. 
This observation trivializes the explicit comparison between $A_{\mathrm{Schur}}$ and $A_{\mathrm{CS}}$. We will review the basic elements of the construction and then focus on the CS interpretation of  the hermiticity structure and module action which emerge in Schur quantization. 

The Seiberg-Witten curve $\Sigma$ that governs the abelian IR physics of the 4D gauge theory is a branched double cover of the curve $C$ defined by the equation $y^2 = \phi_2(z)$ with $\phi_2$
the square of the Seiberg-Witten differential
\be
\label{swform}
\phi_2 = \lambda^2_{\rm SW} = \Bigl(\frac{\Lambda^2}{z} + 2u + \Lambda^2 z\Bigr) \frac{dz^2}{z^2}
\ee
Here $u = {\rm Tr}\phi^2$ denotes the gauge invariant expectation value that parametrizes the Coulomb branch $B_c$. The quadratic differential $\phi_2$ is the expectation value of a 6d BPS operator and 
the singularities of $\phi_2$ at $z=0$ and $z=\infty$ characterize the choice of irregular singularity as having ``rank $\frac12$''.

Upon supersymmetric compactification on a circle, one finds a Coulomb branch of 3d vacua which is a torus fibration over $B_c$. This manifold is hyper-K\"ahler and the fibration over  $B_c$ is holomorphic in complex structure ``$I$''. The functions associated to circle-wrapped BPS line defects are holomorphic in a different complex structure ``$J$''. In this complex structure, the 3d Coulomb branch $\cM$ is a complex symplectic manifold identified as the character variety and $\cA$ is the canonical quantization of its Poisson algebra of holomorphic functions.

The holomorphic functions on the character variety for $C$ are built from the parallel transport of the flat connection and from the Stokes data at the two irregular singularities. The irregular singularities enforce Stokes phenomena for the flat connection with a single Stokes line. We have a non-zero  ``small section'' $s$ and $s'$ at each puncture respectively with the property that the monodromy $M$ around the origin satisfies $s \wedge M \,s \neq 0$ and $s' \wedge M \,s' \neq 0$. There $\wedge$ is the $SL(2)$-invariant antisymmetric inner product.

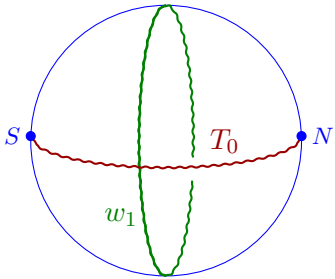
\begin{figure}[t]
\centering
\begin{tikzpicture}[rotate=90]
\begin{scope}[yscale=-1.2,xscale=1.2]
 \tikzset{
    partial ellipse/.style args={#1:#2:#3}{
        insert path={+ (#1:#3) arc (#1:#2:#3)}
    }
}       \draw[blue] [thin] (-2,0) circle (1.5cm);
        \draw[thick,\darkgreen] (-2,0) [partial ellipse=180:360:1.5cm and .3cm];
        \draw[\darkgreen,thick,decorate, decoration={snake, segment length=1.5mm, amplitude=.1mm}](-2.03,0) [partial ellipse=96:-254:1.5cm and .3cm];
        \node[\darkgreen] at (-2.85, -.5) {$w_1$};
        \draw (-1.92,0)[thick,\darkred] [partial ellipse=90:268:.35cm and 1.5cm,decorate, decoration={snake, segment length=1.5mm, amplitude=.1mm}];
        \draw[thin,fill,blue] (-1.95,1.5) circle (.5mm) node [right]{\footnotesize $N$};
        \draw[thin,fill,blue] (-1.95,-1.5) circle (.5mm) node [left]{\footnotesize $S$};      
         \draw[thin,fill,\darkred] (-2.25,.65) circle (0mm) node [above]{$T_0$};
        \end{scope}
\end{tikzpicture}
\caption{The sphere $C$ with two irregular punctures and the line operators $w_1$ and $T_0$ indicated.}
\end{figure}

The classical expectation values of Wilson lines $w_n$ are identified with traces of $M$ in the corresponding irreps of $SL(2)$. The elementary operators with magnetic charge $2$ are defined as invariant ``cross-ratios'' of small sections transported around certain paths on $C$. If we normalize 
\begin{equation}
    s \wedge Ms =s' \wedge Ms' =1 
\end{equation}  
the classical expectation value of $\tH_a$ is identified with $s'\wedge M^a s$. As $s$ and $s'$ are defined modulo sign, we should only really consider products of an even number of $\tH_a$'s. 

Recall that $M + M^{-1} = \Tr M$ for any $SL(2)$ matrix. Then the classical skein
relation $w_1 \tH_a = \tH_{a+1} + \tH_{a-1}$ is obvious. Also
\begin{align}
    \tH_1 \tH_{-1} -\tH_0^2 & = (s'\wedge Ms) (M s'\wedge s)- (s'\wedge s)(M s' \wedge Ms)\nonumber \\[2mm] & = - (s'\wedge M s')(s \wedge M s)=-1\, ,
\end{align}
etcetera. With some minimal extra work, one can derive the classical limit of all the skein algebra relations \eqref{wthalgebra}-\eqref{wthalgebrat}.

The twist $\rho^2$ which appears in the Schur trace is an important subtlety in the story. The automorphism $\rho^2$ encodes the effect of a $2\pi$ rotation for the anomalous $U(1)_r$ symmetry of the theory, which occurs because the HT twisted theory mixes $U(1)_r$ rotations and rotations of the topological plane. In the current context this is the $U(1)$ symmetry which rotates the SW differential and is broken by the phase of $\Lambda$. As the phase of $\Lambda$ rotates, the position of the Stokes rays at the irregular singularities rotates as well and $\rho^2$ induces a full shift of the Stokes sectors around each irregular singularity. In the case at hand, the result is a $4 \pi$ rotation  of the single Stokes ray around each of the irregular puncture, $\rho^2(\tH_a) = \tH_{a-4}$ \cite{Gaiotto:2010be}.

The action of $\rho$ should be associated to a $2 \pi$ rotation. In order to lift the action to odd products of $\tH_a$'s we need to lift the sign ambiguity. Positivity in complex Schur quantization 
gives $\rho(\tH_a) = - \tH_{a-2}$. The sign is immaterial when we restrict to $\cA$.

An useful perspective on $\rho$ is that we start from an holomorphic function $a$ in complex structure $J$ and continuously deform it into 
an holomorphic function in complex structure $\cos \phi J + \sin \phi K$,
which gives the same character variety with rotated Stokes lines. We land at $-J$ with an anti-holomorphic function which we can conjugate to 
obtain $\rho(a)$. 

\smallskip

\subsection{Reality structures and modules}

\vspace{-1mm}

We now specify the geometric implementation of the reality conditions that define the $*$ structure on the algebra $\cA$.

Real Schur quantization is connected to the quantization of $G_\bC$ Chern-Simons theory by a duality chain starting with the class ${\cal S}$ construction of the 4d gauge theory as a 6d SCFT compactified on $C$. This class ${\cal S}$ description relates supersymmetric correlators in gauge theory to correlators in the CS theory on a three manifold $M_3$. Besides the well-known identification $\cA_{\mathrm{Schur}} = \cA_{\mathrm{CS}}$ between the operator algebra on both sides \cite{Gaiotto:2024tpl}, the connection involves two pieces of data which are related in a somewhat subtle way:
\begin{itemize}
    \item The module ${\cal M}_{\mathrm{Schur}}$ matches a skein module ${\cal M}_{\mathrm{CS}}$ associated to the auxiliary three-manifold $M_3$ with boundary $C$. The class $S$ dictionary associates $M_3$ to the Neumann boundary condition for SW theory on $\bR^3 \times \bR^+$.  
    \item The Hilbert space ${\cal H}_{\mathrm{Schur}}$ matches the Hilbert space ${\cal H}_{\mathrm{CS}}$ of $G_\bC$ Chern-Simons theory on the quotient surface $C_\bR = C/\tau$, i.e. $C$ is the orientation cover of $C_\bR$. The manifold $M_3$ is a smoothening of the quotient $(C \times [-1,1])/\bZ_2$ defined by simultaneous reflection of the two factors.
\end{itemize}
The existence of a map ${\cal M}_{\mathrm{CS}} \to {\cal H}_{\mathrm{CS}}$ compatible with the $\cA_{\mathrm{CS}}$ action on both is not immediately obvious \cite{Gaiotto:2024osr}, but it can be motivated classically as follows. 

The image of ${\cal M}_{\mathrm{CS}}$ is expected to consist of boundary states built from a space-like $SU(2)$ boundary condition (decorated by boundary Wilson lines) that requires the complex connection on $C_\bR$ to be unitary on the boundary $C$. We can lift that to a connection on $C$ which is both unitary and invariant under $\tau$.
The manifold $M_3$ imposes invariance under the combination of the two transformations, so an unitary connection on $C_\bR$
is the same as a a connection which extends to ${\cal M}_{\mathrm{CS}}$ when lifted to $C$.

We now propose a specific geometric implementation of the map $\tau$ defining the $*$-structure on $\cA$. As $\tau$ inverts the algebra operation, 
it must also reverse the orientation on $C$. A simple physical argument is that the 6d theory is chiral and $\tau$ involves a reflection in 4d. 
It must thus reflect $C$ as well. There are several options:

\begin{enumerate}
    \item A reflection which fixes the location of the irregular singularities, so that the quotient $C_\bR$ of $C$ by the reflection becomes a disk with two boundary irregular singularities.
    \item A reflection which permutes the two irregular singularities and fixes the equator, so that the quotient $C_\bR$  of $C$ by the reflection becomes a disk with a single irregular singularity at the center. 
    \item A reflection which permutes the two irregular singularities and has no fixed points, so that the quotient $C_\bR$ of $C$ by the reflection becomes a cross-cap with a single irregular singularity at the center. 
\end{enumerate}
On top of that, we would also have to select how the reflection acts on the gauge bundle at the fixed locus of the reflection, i.e. the gauge group at the boundary of the quotient geometry.

Some of the options appear immediately excluded by the form of $\tau$. For example, option 3. acts on 't Hooft lines as $\tH_a \to \tH_{1-a}$ and does not fix any of them. The first two options are a bit more difficult to distinguish from each other, as the combination of the two geometric transformations appears to act trivially on $\cA$. 

In the following we adopt the second option 2. This choice is motivated by a comparison between the module ${\cal M}_{\rm CS}$ associated to Wilson line operators on $M_3$ and the module ${\cal M}_{\rm Schur}$ constructed by the real Schur quantization procedure outlined in section 2.4. The gauge group at the boundary of $C_\bR$ determines the classical spectrum of the Wilson line operators. Based on the quantum spectrum that we are trying to match, we take the boundary gauge group to be $SU(2)$.

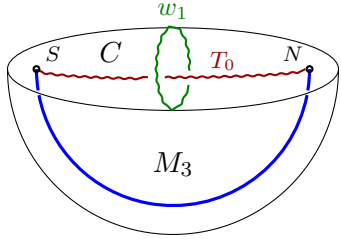
\begin{figure}[t]
\centering
{\begin{tikzpicture}[rotate=0,xscale=1.1,yscale=1.1] 
\tikzset{
    partial ellipse/.style args={#1:#2:#3}{
        insert path={+ (#1:#3) arc (#1:#2:#3)}
    }
}   
\draw[black] (0,0) [partial ellipse=-180:180:2cm and .5cm];
\draw[blue, very thick] [partial ellipse=11:169:1.65cm and -1.65cm];
\draw[blue, very thick] [partial ellipse=0:9:1.65cm and -1.65cm];
\draw[blue, very thick] [partial ellipse=171:180:1.65cm and -1.65cm];
\draw [partial ellipse=0:180:2cm and -2cm];
\draw[\darkgreen,thick,decorate, decoration={snake, segment length=1.5mm, amplitude=.1mm}](0,-.02) [partial ellipse=0:340:.2cm and .5cm];
        \draw (0,-.05)[thick,rotate=90,\darkred] [partial ellipse=95:168:.1cm and 1.7cm,decorate, decoration={snake, segment length=1.5mm, amplitude=.1mm}];
        \draw (0,.1)[thick,rotate=90,\darkred] [partial ellipse=180:270:.1cm and 1.7cm,decorate, decoration={snake, segment length=1.5mm, amplitude=.1mm}]; 
\draw[\darkred] (.6,.1) node {\mbox{\footnotesize $T_0$}};
 \draw[\darkgreen] (0,.7) node {\mbox{\small $w_1$}};
\draw[thick] (-1.65,0) circle (.35mm);
\draw[thick] (1.65,0) circle (.35mm);
\draw (1.45,.18)  node {\raisebox{0mm}{\scriptsize $N$}};
\draw (-1.45,.18)  node {\raisebox{0mm}{\scriptsize $S$}};
\draw (0,-1.15)  node {\raisebox{0mm}{$M_3$}};
\draw (-.75,.2)  node {\raisebox{0mm}{$C$}};
\end{tikzpicture}}
\caption{The half ball $M_3$ with the bulk line defect ending at the irregular punctures on boundary $C$, with the line operators $w_1$ and $\tH_0$ indicated.}
\end{figure}

Next, we want to determine the class ${\cal S}$ interpretation of Neumann b.c. The 6d theory has no known boundary conditions. Instead, one can consider a 3d manifold $M_3$ with a cylindrical end of cross-section $C$, with some defect network connected to the two irregular singularities of $C$. 
This produces a boundary condition for the 4d theory. 
Upon compactification on a circle, this defines a sub-manifold ${\mathcal L}$ of the space of flat connections on $C$ consisting of flat connections which extend to $M_3$. This controls the behaviour of line defects brought to the boundary. Boundary line defects are described by skeins in $M_3$.

The 3d manifold $M_3$ must enforce classical constraints $\tH_1=0$, $\tH_{2+a} = -w_a$ and $\tH_{-a} = w_a$. The vanishing of $\tH_1$ suggests a condition $s' = Ms$. This implies 
\begin{align}
    \tH_{2+a} &= Ms \wedge M^{a+2} s = s \wedge M^{a+1} s= w_1 (s \wedge M^{a} s)- s \wedge M^{a-1} s \nonumber\\[1.75mm]
    \tH_{-a} &= Ms \wedge M^{-a} s = s \wedge M^{-a-1} s= w_1 (s \wedge M^{-a} s)- s \wedge M^{-a+1} s
\end{align}
which indeed imply the desired constraints. 

We can thus define a convenient $M_3$ as a three-ball 
with boundary $C$ and an irregular defect which joins the two singularities at the boundary with a $2 \pi$ twist 
of the local Stokes line in the plane transverse to the defect. This insures the desired relation between $s$ and $s'$. The $w_n$ holonomies on $C$ are not contractible in $M_3$ and give boundary lines identified with $w^\partial_n$. 

We then identify the boundary Wilson lines $w_n^\partial$ in the Schur context with elements of the ``Skein module'' of quantized holonomies on this $M_3$ geometry. We expect this Skein module to reproduce the module ${\cal M}$.  

Motivated by our (brief) discussion of brane quantization, we would like to give a slightly different interpretation of $M_3$: flat connections on $M_3$ are the same as flat connections on $C$ which are invariant under the action of $\rho$ followed by a reflection $\tau$ exchanging the two irregular singularities and acting as the identity on the equator. This makes it likely that hyper-Kahler rotation of the manifold ${\mathcal L}$ gives a real locus in $\cM$ which consists of complex connections which are invariant under $\tau$ combined with complex conjugation. This is the same as the space of complex connections on the disk $\bC_\bR$ with an irregular singularity at the origin and a $SU(2)$ structure group at the boundary. This is our candidate $\cM_\bR$.

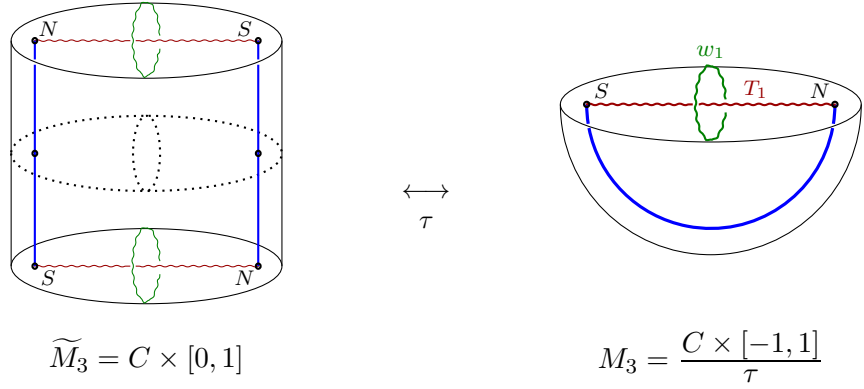
\begin{figure}[t]
\centering
\begin{tikzpicture}[rotate=0,xscale=.9,yscale=1] 
\tikzset{
    partial ellipse/.style args={#1:#2:#3}{
        insert path={+ (#1:#3) arc (#1:#2:#3)}
    }
}   
\draw[black] (0,-1.5) [partial ellipse=-180:180:2cm and .5cm];
\draw[black] (0,1.5) [partial ellipse=-180:180:2cm and .5cm];
\draw[dotted,thick] (0,0) [partial ellipse=-180:180:2cm and .5cm];
\draw[dotted,thick] (0,0) [partial ellipse=-180:180:.2cm and .5cm];
\draw[\darkred,decorate, decoration={snake, segment length=1.5mm, amplitude=.1mm}] (-1.66,1.5)--(-0.24,1.5);
\draw[\darkred,decorate, decoration={snake, segment length=1.5mm, amplitude=.1mm}] (1.66,1.5)--(-0.16,1.5);
\draw[\darkgreen,decorate, decoration={snake, segment length=1.5mm, amplitude=.1mm}](0,1.48) [partial ellipse=10:350:.2cm and .5cm];
\draw[\darkred,decorate, decoration={snake, segment length=1.5mm, amplitude=.1mm}] (-1.66,-1.5)--(-0.24,-1.5);
\draw[\darkred,decorate, decoration={snake, segment length=1.5mm, amplitude=.1mm}] (1.66,-1.5)--(-0.16,-1.5);
\draw[\darkgreen,decorate, decoration={snake, segment length=1.5mm, amplitude=.1mm}](0,-1.52) [partial ellipse=10:350:.2cm and .5cm];
\draw[black] (2,-1.5) -- (2,1.5);
\draw[black] (-2,-1.5) -- (-2,1.5);
\draw[blue, thick] (1.65,-1.5) -- (1.65,1.18) ;
\draw[blue,thick] (1.65,1.25) -- (1.65,1.5) ;
\draw[blue,thick] (-1.65,-1.5) -- (-1.65,1.18) ;
\draw[blue,thick] (-1.65,1.25) -- (-1.65,1.5) ;
\draw[black,thick] (-1.65,-1.5) circle (.35mm);
\draw[black,thick] (1.65,0) circle (.35mm);
\draw[black,thick] (-1.65,0) circle (.35mm);
\draw[black,thick] (1.65,-1.5) circle (.35mm);
\draw[black,thick] (-1.65,1.5) circle (.35mm);
\draw[black,thick] (1.65,1.5) circle (.35mm);
\draw[black] (1.45,-1.66)  node {\raisebox{0mm}{\scriptsize $N$}};
\draw[black] (1.45,1.66)  node {\raisebox{0mm}{\scriptsize $S$}};
\draw[black] (-1.45,-1.66)  node {\raisebox{0mm}{\scriptsize $S$}};
\draw[black] (-1.45,1.66)  node {\raisebox{0mm}{\scriptsize $N$}};
\draw[black] (0,-2.7) node {\mbox{$
    \widetilde{M}_3 = {C \times [0,1]}$}};
\end{tikzpicture}
~~~~~~~~~~~\raisebox{2.25cm}{{\Large ${\longleftrightarrow}\atop {\tau}$}}~~~~~~~~~~
\raisebox{-1mm}{\begin{tikzpicture}[rotate=0,xscale=1,yscale=1] 
\tikzset{
    partial ellipse/.style args={#1:#2:#3}{
        insert path={+ (#1:#3) arc (#1:#2:#3)}
    }
}   
\draw[black] (0,0) [partial ellipse=-180:180:2cm and .5cm];
\draw[blue, very thick] [partial ellipse=11:169:1.65cm and -1.65cm];
\draw[blue, very thick] [partial ellipse=0:9:1.65cm and -1.65cm];
\draw[blue, very thick] [partial ellipse=171:180:1.65cm and -1.65cm];
\draw [partial ellipse=0:180:2cm and -2cm];
\draw[\darkred,thick,decorate, decoration={snake, segment length=1.5mm, amplitude=.1mm}] (-1.66,0)--(-0.24,0);
\draw[\darkred,thick,decorate, decoration={snake, segment length=1.5mm, amplitude=.1mm}] (1.66,0)--(-0.16,0);
\draw[\darkgreen,thick,decorate, decoration={snake, segment length=1.5mm, amplitude=.1mm}](0,-.02) [partial ellipse=10:350:.2cm and .5cm];
\draw[\darkred] (.6,.2) node {\mbox{\scriptsize $T_1$}};
 \draw[\darkgreen] (0,.7) node {\mbox{\footnotesize $w_1$}};
\draw[thick] (-1.65,0) circle (.35mm);
\draw[thick] (1.65,0) circle (.35mm);
\draw (1.45,.18)  node {\raisebox{0mm}{\scriptsize $N$}};
\draw (-1.45,.18)  node {\raisebox{0mm}{\scriptsize $S$}};
\draw (0,-3.35) node { \raisebox{1mm}{$M_3 =$}~\raisebox{1mm}{$\frac{\mbox{$C\times [-1,1]$}}{\raisebox{-.5mm}{$\tau$}}$}};
\end{tikzpicture}}
\caption{The space $M_3$ can be presented as a quotient, up to ``smoothening'' a $\bR^2/\bZ_2$ conical singularity. If we quotient $C$ by $\tau$, we obtain the real surface $C_\bR$. We refer to \cite{Gaiotto:2024osr} for the description of the full 4-d geometry relating these quotients and brane quantization of the character variety for $C_\bR$.}\label{fig:m3}
\end{figure}

It is easy to see that the classical version of $w_n$ and $\tH_a$ 
restricted to $\cM_\bR$ satisfy the expected reality condition. 
Indeed, $w_n$ can be computed along the boundary of the disk 
as an $SU(2)$ holonomy, while $\tH_a$ is computed by transporting $s$ and its complex conjugate in opposite directions around the origin and then contracting them at the boundary with the canonical $SU(2)$-invariant Hermitean pairing there. 

\smallskip

In conclusion, we have found the following chain of identifications:
\begin{itemize}
    \item Neumann b.c. for pure $SU(2)$ SW theory have a geometric class $S$ interpretation as a 3d manifold $M_3$ with boundary $C$. 
    \item The manifold $M_3$ is topologically equivalent to a quotient of $C \times [-1,1]$ by a simultaneous reflection $\tau$ acting on $C$ and $[0,1]$, see Figure \ref{fig:m3}. Following \cite{Gaiotto:2024osr}, this indicates that real Schur quantization based on Neumann b.c. will lead to the quantization of $SL(2,\mathbb{C})$ CS on $C/\tau \equiv C_\bR$ with specific boundary conditions on the disk: the connection restricted to the boundary should be unitary.
\end{itemize}

\subsection{Real Schur quantization of $SL(2,\mathbb{C})$ CS theory on $C_\bR$.}

We now review how the output of real Schur quantization is identified with a quantization of the $SL(2,\bC)$ CS theory on the disk $C_\bR$, with a boundary condition which forces the connection to be unitary. This boundary condition is well-defined and is topological, as the CS action with pure imaginary level restricted to unitary connections vanishes. 

The CS action equips the character variety ${\cal M}(SL(2),C)$ with a symplectic form
\be
\Omega = \frac{i\kappa}{4\pi} \int_C {\rm tr}( \delta A \wedge \delta A).
\ee
Upon quantization, the $SL(2)$ holonomies generate a non-commutative Skein algebra Sk$_\fq(C,{\mathfrak sl}_2)$  \eqref{wthalgebra}-\eqref{wthalgebrat} with $q = e^{-\frac{2\pi}{\kappa}}$.\footnote{A small subtlety is that one could consider a variant of the theory employing a spin-$SL(2,\bC)$ gauge group, i.e. require spin $1/2$ Wilson lines to behave a worldlines of space-time spinors. This modifies the ``skein relations'' 
in such a way that $q$ is negative instead of positive.}
Enforcing the reality conditions, the phase space of the system is precisely $\cM_\bR$ and the algebra of observables is naturally quantized as a skein algebra on $\bC_\bR$, i.e. the skein algebra $\cA$ on $C$ equipped with an Hermiticity condition $\tau$:
\be
 w_n^\dagger = w_n \qquad\qquad \tH_a^\dagger = \tH_{-a}\\[1mm]
\ee 
Classically, the $w_n$ can be brought to the boundary of $C_\bR$ and are traces of an element in $SU(2)$.
Real Schur quantization gives us precisely an Hilbert space $\cH = L^2(S^1/\bZ_2,d\mu)$ equipped with an unitary representation of the desired operator algebra, such that the spectrum of $w_n$ consists of traces of an element in $SU(2)$. 

Recall that the representation is essentially unique if we assume the existence of a state $|0\rangle \equiv 1^\partial$ which satisfies the relations 
\begin{align}  
& T_{1} |0 \ra = 0 \cr & T_0| 0 \ra = w_1|0\ra = |0\ra
\end{align}
Indeed, consider the collection of states 
\begin{equation}
    |n;w\rangle \equiv w_n|0\rangle \, .
\end{equation}
Under the action of $w_1$, $w_n|0\rangle$ and $\fq^{\frac{1+n}{2}} \tH_{-n} |0\rangle$ satisfy the same three-term relation. The initial conditions at $n=-1$ and $n=0$ 
for the recursion determined by the relation are the same, so these states coincide for all positive $n$ as well. The sequence $-\fq^{-\frac{1+n}{2}} \tH_{n+2} |0\rangle$ also coincides with $w_n|0\rangle$ by the same token. This gives the broader set of relations 
\begin{align}  
& T_{1} |0 \ra = 0, \qquad T_0| 0 \ra = w_1|0\ra = |0\ra\notag \\[1mm]
& w_n|0\rangle =\fq^{\frac{1+n}{2}} \tH_{-n} |0\rangle = -\fq^{-\frac{1+n}{2}} \tH_{n+2} |0\rangle \, ,
\end{align}
which are sufficient to define the abstract $\cA$ module ${\cal M}$ with linear basis $|n;w\rangle$. Indeed, the action of $w_m$ on  $|n;w\rangle$ follows from the expansion of $w_m w_n$ into a sum of $w_i$'s and the action of $\tH_a$ follows from the expansion of $\tH_a w_n$ into a sum of $\tH_{a+i}$'s.

The state $|0\rangle$ is interpreted in \cite{Gaiotto:2024osr} as being produced by a space-like boundary condition where the gauge connection is also reduced to $SU(2)$, i.e. a configuration where the disk $C_\bR$ itself shrinks smoothly to a point where 
the irregular singularity somehow also ends. The states $|n;w\rangle$ are then produced by an extra $SU(2)$ holonomy defect placed at the space-like boundary. 
Any quantization of the system which admits such boundary states should thus coincide with the real Schur quantization, up to completely decoupled super-selection sectors. 
It would be interesting to explore this perspective further. 

The dictionary to Schur indices becomes transparent in a different spectral presentation of $\cH$ which diagonalizes $w_1$. Obviously, $w_1$ acts on $\fq^{-\frac{a}{2}} \tH_a |0\rangle$ as a very simple difference operator. We can build formal eigenfunctions of $w_1$
\begin{equation}
    \sum_{a\in \bZ} e^{i a \theta} \fq^{\frac{a}{2}} \tH_{1-a} |0\rangle
\end{equation}
with eigenvalue $2\cos\theta$. Or with a slightly better normalization, we define
\begin{equation}
 |\theta\ra =   \sum_{n \geq 0} \frac{\sin (n+1) \theta}{\sin \theta} \, |n;w\rangle = \sum_{n \geq 0} \chi_n(e^{i \theta}) \, |n;w\rangle
\end{equation}
 These states form an orthonormal basis and are distributional, {\it i.e.} they are delta function normalized with respect to the spectral measure $d\mu$. 

We are now ready to make contact with the DSSYK model.

\smallskip

\section{Duality between Schur quantization and double scaled SYK}

\vspace{-1mm}

The purpose of this section is to show that the algebra of observables in the double-scaled SYK model is isomorphic to the Skein algebra $\cA$ introduced in the previous Sections. 

\subsection{Double scaled SYK Hilbert space}
\vspace{-1mm}
In the double scaling limit, the computation of moments of the SYK Hamiltonian ${\bf H}$ reduces to the counting all chord diagrams weighted by of factor $\mfq$ for each intersection \cite{Berkooz:2018jqr}. For the partition function, one only considers the Hamiltonian chords. To perform this summation, it is useful to slice open a chord diagram and introduce an auxiliary Hilbert space spanned by basis states $|n\ra$ labeled by the chord number $n$ \cite{Berkooz:2018jqr,Lin:2023trc}.  The Hamiltonian acts on the basis states $|n\ra$  as  \cite{Berkooz:2018jqr}
\be
 \label{hamrule}
{\bf H} | n \ra = | n + 1\ra + [n]_\mfq \, |n -1\ra, \qquad
[n]_\mfq \equiv\frac{1-\mfq^{n}}{1-\mfq}.
\ee
This rule reflects that acting with ${\bf H}$ can either add a new chord or remove an existing chord. A new chord is prescribed to have no intersections while a chord that is removed can only do so by intersecting all chords between itself and the new Hamiltonian insertion. 

The action of ${\bf H}$ on the chord number basis is conveniently summarized in terms of $\mfq$-deformed oscillators $\mfa^\dag$ and $\mfa$ defined via
\be
\mfa^\dag | n\ra = \sqrt{[n+1]\;}{\!\!}_{\mfq}\,|n+1\ra, \qquad \quad \mfa | n \ra = \sqrt{[n]\;}{\!\!}_{\mfq}\, |n-1\ra
\ee
Let $\mfn$ be the number operator $\mfn | n \ra = n |n\ra$. Then
\be
\mfa^\dag \mfa  =[\mfn]_\mfq,\quad \quad\ \mfa\,\mfa^\dag  =  [\mfn + 1]_\mfq, \qquad \ [\mfa,\mfa^\dag]_\mfq  = 1.
\ee
We can write
\be
\label{haa}
{\bf H} = \mfa^\dag + \mfa \, 
\ee
In what follows, we will focus one the operator algebra generated by the Hamiltonian ${\bf H}$ and the operator $\mfq^{-\mfns/2}$ with $\mfn$ the number operator that counts the number of chords.\footnote{
The q-deformed raising and lowering operators $\mfa$ and $\mfa^\dag$ and the number operator $\mfns$ are coarse grained collective observables. In a two-sided microscopic Hilbert space ${\cal H}_L\otimes {\cal H}_R$, spanned by all states obtained by acting with SYK operators on the TFD state, the total chord number operator $\mfns$ can be expressed in terms of the fundamental majorana oscillators  as 
\cite{Lin:2023trc}
\be
\mfn = \frac{1}{2p}\sum_{i=1}^{N} \bigl(1 + \imath \psi^L_i \psi^R_i \bigr)
\ee
One can then obtain a microscopic expression for the $\mfq$ deformed oscillators $\mfa$ and $\mfa^\dag$ in terms of the fundamental majorana oscillators by using the above identity for $\mfn$ and the relations \eqref{haa} and $[\mfn, \HH]  = \mfa^\dag - \mfa$.}

\def\cE{{\cal E}}

\subsection{Skein algebra from double scaled SYK}
We will now show that the operator algebra generated by ${\bf H}$ and $\fq^{-\mfn-\frac12}$ gives the algebra $\cA'$.
We introduce the notation 
\bea
\label{raiselower}
\yp =   
\, \sqrt{1-q^2}\, \mfa^\dag
\qquad\;  & &  \ \ \ \, \ym =  \sqrt{1-\fq^2}\, 
\mfa \\[2mm]
w_1 \, = \sqrt{1-\fq^2}\, {\HH} \qquad  & & \  \ \ {\tH}_0  =\, \fq^{-\mfn-\frac12}\,
\eea
The operator  $\tH_0$ is diagonal on the chord basis of states $|n\rangle \in \c\tH_\fq$. The operators $y_\pm$ and $T_0$ act on each other via
\begin{align}
\yp \tH_0  & = q \, \tH_0\,  \yp, \qquad \qquad \quad\, \ym \yp  = 1 - q \, \tH_0^{-2},\\[.5mm]
\ym \tH_0  & = q^{-1} \tH_0 \, \ym, \qquad \qquad \    \yp \ym = 1 -\fq^{-1} \tH_0^{-2} 
\end{align}
The action of $w_1 = \yp + \ym$ in the chord representation is expressed in terms of a $\fq$-deformed Weyl algebra $W_\fq$.

We define $\cA'$ as the sub-algebra of $W_\fq$ generated from $\tH_0$ and $w_1$. We now claim that, starting from $\tH_0$ and taking repeated $\fq$-commutators with $w_1$, we obtain an infinite family of operators $\tH_a \in \cA'$ labeled by an integer $a$ via the recursion relations:
\begin{align}
    w_1 \tH_a &= \fq^{-\frac12} \tH_{a+1} + \fq^{\frac12} \tH_{a-1} \cr
    \tH_a w_1 &= \fq^{\frac12} \tH_{a+1} + \fq^{-\frac12} \tH_{a-1}
\end{align}
The proof is not difficult. We first define
\begin{align}
    \tH_{1} &= \frac{\fq^{-\frac12} w_1 \tH_0 -\fq^{\frac12}\tH_0 w_1}{\fq^{-1}-\fq}=\fq^{\frac12} \ym \tH_0  \cr
    \tH_{-1} &= \frac{\fq^{-\frac12} \tH_0 w_1 -\fq^{\frac12}w_1 \tH_0 }{\fq^{-1}-\fq} = \fq^{-\frac12} \yp \tH_0 
\end{align}
and then use the first of the two recursions to define the $\tH_a$, e.g. for positive $a$:
\begin{equation}
    \tH_{a+1} = \fq^{\frac12} w_1 \tH_a - \fq \tH_{a-1}
\end{equation}
Then one can write a three step recursion for the expressions $\tH_a w_1 - \fq^{\frac12} \tH_{a+1} - \fq^{-\frac12} \tH_{a-1}$ and verify they vanish for $a=0$ and $a=1$. We compute
\begin{align}
    \tH_{2} &= \fq \ym^2 \tH_0 - \tH_0^{-1} \notag\\[.2mm]
    \tH_{1} &= \fq^{\frac12}\ym \tH_0  \notag\\[.5mm]
    \tH_0 &= \tH_0 \\
    \tH_{-1} &= \fq^{\frac12} \tH_0\, \yp \notag\\[.5mm]
    \tH_{-2} &= \fq \tH_0 \,\yp^2 - \tH_0^{-1} \cr
    & \cdots \nonumber
\end{align}
We observe the Hermiticity condition $\tH_a^\dagger = \tH_{-a}$. We also find the remarkable relations 
\begin{align}
    \tH_a \tH_{a+1} &= \fq \tH_{a+1} \tH_a \notag\\[.5mm]
    \tH_{a-1} \tH_{a+1} &= -1 + \fq \tH_{a}^2 \\[.5mm]
    \tH_{a+1} \tH_{a-1} &= -1 + \fq^{-1} \tH_a^2 \, .\notag
\end{align}
We can verify these identities for small $a$ and then recur by judicious applications of $w_1$ from the left and from the right. 
Finally, we define elements $D_{m,a} \in \cA'$ and $w_n$ via 
\begin{align}
    & D_{n+m,an+am+m} \equiv \fq^{-\frac{n m}{2}}\tH_a^n \tH_{a+1}^m \, .\\[2.5mm]
    &\qquad w_{n+1} = w_1 w_n - w_{n-1}
\end{align}
The above relations are sufficient to show that the generators form a linear basis (over $\bZ[\fq^{\pm \frac12}]$) of the algebra $\cA'$: any product can be expanded into a finite linear combination of generators. 

We can now observe which SYK operators correspond to the $\cA$ sub-algebra. We have $q \tH_0^2 = q^{-2\mfn}$, corresponding to the ${\cal O}_{-1} {\cal O}_{-1}$ insertion. We also have relations such as 
\begin{align}
   \fq^{-1} w_1 (\tH^2_a) - \fq (\tH_a^2) w_1\, =\hspace{7mm} &\hspace{-5mm} (\fq^{-2} - \fq^{2}) (\fq^{-\frac12}\tH_a \tH_{a+1} )\notag\\[-2mm]\\[-2mm]\notag
   \fq^{-1} w_1 (\fq^{-\frac12}\tH_a \tH_{a+1}) - \fq (\fq^{-\frac12}\tH_a \tH_{a+1}) w_1 &\,=\,  (\fq^{-2}-\fq^2) \tH_{a+1}^2 +\fq-\fq^{-1}
\end{align}
which allow one to generate all elements of $\cA$ out of $\fq \tH_0^2$ and $w_1$. If we work with $\fq \tH_a^2$ and $(\fq^{-\frac12}\tH_a \tH_{a+1} )$, the relations only employ integer powers of $\mfq$, so even the choice of square root $\fq$ drops out. 
The $*$-structure on the algebra is also manifestly $w_1^\dagger =w_1$ and $\tH_a^\dagger = \tH_{-a}$.

\subsection{Embedding the Skein module in $\cH_{\rm DSSYK}$}

For completeness, we can describe a direct unitary transformation between the DSSYK Hilbert space and the auxiliary Hilbert space $L^2([0,\pi],d\mu)$ encountered in the description of $\cH_{\mathrm{Schur}}$. 
We have already established that the algebraic structure is sufficiently rigid to fix the inner product on $\cM$, so it must coincide with the inner product in $\cH$. As a payoff, we recover the known integral expression of $\langle 0|w_n|0\rangle$ as $I\!\!I_{w_n}(\fq)$.

The vacuum $|0\rangle$ manifestly satisfies 
\begin{align}
    T_2 |0\rangle = - \fq^{\frac12} |0\rangle, \qquad\
    T_1 |0\rangle = 0,\qquad\
    T_0 |0\rangle = \fq ^{-\frac12} |0\rangle
\end{align}
Acting with $w_1$, we can recover the remaining relations expected for $1^\partial$ in $\cM$. E.g. 
\begin{align}
   \fq^{-\frac12} T_3 |0\rangle = - \fq^{\frac12} w_1|0\rangle, \qquad \ \ 
   \fq^{\frac12} T_{-1} |0\rangle = \fq ^{-\frac12} w_1|0\rangle,
\end{align}
etcetera. These relations are sufficient to identify $w_a^\partial$ with $w_a |0\rangle$. This is manifestly a linear combination $|n\rangle$ with $n \leq a$, with a non-zero coefficient for $|a\rangle$. The relation between the $w_a^\partial$ and $|n\rangle$ states is thus triangular and invertible. 

As we discussed above, $|n\rangle$ can be expanded as a linear combination of $w_a|0\rangle$ for $a\leq n$. If we write $w_a = \chi_a(v)$, we can write 
\begin{equation}
    |n\rangle = \frac{1}{\sqrt{(1-\mfq)^n [n]_\mfq!}}\; P_n(v) |0\rangle
\end{equation}
for some palindromic Laurent polynomial $P_n(v)$ of degree $v$. This imposes the recursion 
\begin{equation}
    (v+v^{-1})P_n(v)= P_{n+1}(v) + (1-\mfq^n) P_{n-1}(v)
\end{equation}
which can be recognized as the recursion relation for the $\mfq$-Hermite polynomials. This recursion formula and the polynomials $P_n(\zeta)$ play a key role in the exact solution of the DSSYK model.

The $\mfq$-Hermite polynomials can be collected into a generating function 
\begin{equation}
\label{genfun}
    P(v;\sigma) = \sum_{n=0}^\infty \, \frac{\sigma^n}{
    (1-\mfq)^n [n]_{\mfq}!}\, P_n(v)
\end{equation}
with $[n]_{\mfq}!=[1]_\mfq \cdots [n]_{\mfq}$ the $\mfq$-factorial. The recursion formula for the generating function
\begin{equation}
    P(v;\fq^2 \sigma)  = (1-\sigma v)(1-\sigma v^{-1})P(v;\sigma)
\end{equation}
is solved via
\begin{equation}
\label{genpfun}
	P(v;\sigma) = \frac{1}{(v \sigma;\mfq)_\infty (v^{-1} \sigma;\mfq)_\infty}
\end{equation}
The above formula is the $\mfq$-deformation of the familiar gaussian generating function of the classical Hermite polynomials.

From the explicit expression \eqref{genpfun} of the generating function $P(v;\sigma)$, we can derive two crucial difference equations
\begin{align}
&\ \ \fq^{-n} (v-v^{-1}) P_n(v)\, =\, v P_n(\fq^{-1} v)-v^{-1} P_n(\fq v) \, \\[2.5mm]
	& (\fq^n - \fq^{-n})  (v-v^{-1}) P_{n-1}(v) =P_n(\fq v)-P_n(\fq^{-1} v)\, ,
\end{align}
In these equations, we recognize the the quantum Fenchen-Nielsen expressions for the 't Hooft operators $\tH_0$ and $\tH_{1}$ on the right-hand side up to a factor of $v-v^{-1}$. \footnote{The kernel $P(v;\sigma)$ can be recognized as the half-index of the ``RG flow interface'' 3d theory from \cite{Dimofte:2013lba}, which indeed transforms 
the 't Hooft lines corresponding to $\tH_0$ and $\tH_{1}$ into simpler line defects in an Abelian gauge theory. The next Section will illustrate the role of 3d interfaces in real Schur quantization.}

\def\cE{{\cal E}}

\section{Correlators with matter operators}

\vspace{-1mm}

To make the dictionary between the observables in Schur quantization and DSSYK more complete, we also need to be able to include correlation functions involving arbitrary combinations of local DSSYK operators ${\cal O}_\Delta$. Here we make some preliminary comments about how these more general matter operators may be included into our story.

\subsection{Matter two-point function}
\vspace{-1mm}

As a first example we consider the DSSYK two-point function of two scaling operators
\be
\label{twopt}
G_\Delta(\beta_1,\beta_2) = {\rm Tr}\Bigl(
e^{-\beta_1{\bf H}}{\cal O}_\Delta  e^{-\beta_2{\bf H}}\,{\cal O}_\Delta \Bigr) \, =\,
\begin{tikzpicture}[scale=0.64, rotate=0, baseline={([yshift=-.07cm]current bounding box.center)}]
\draw[thick] (0,0) circle (1.5); \draw[thick,\darkblue] (1.5,-.0) -- (0,-.0); 
\draw[thick,\darkblue] (-1.5,-.0) -- (0,-.0);
\draw[fill=blue] (-1.5,.0) circle (0.1); 
\draw[fill=blue] (1.5,0) circle (0.1);
\draw (0,1.95) node {\footnotesize $\beta_2$}; 
\draw (0.05,-1.95) node {\footnotesize $\beta_1$}; 
\draw (1.95,0) node {\footnotesize \textcolor{\darkgreen}{$\Delta$}};
\draw (-1.95,0) node {\footnotesize \textcolor{\darkgreen}{$\Delta$}};
\end{tikzpicture}
\ee
In terms of the chord basis $|n\ra$, the matter chord between two local operators ${\cal O}_\Delta$ with scaling dimension $\Delta$ (that is, given by a sum of products of $p' = p \Delta$ Majorana oscillators) is~represented via the operator $\mfq^{\mfns\Delta} = q^{2\mfns\Delta}$ 
with $\mfn$ the chord number operator that counts the intersections between the matter and Hamiltonian chords. 

Alternatively, we can 
write the two point function \eqref{twopt} as the overlap
\begin{align}
\label{gtwotwo}
G_\Delta(\beta_1,\beta_2) 
& =    \langle 0,0| e^{-\beta_1 {\bf H}_L-\beta_2 {\bf H}_R}  | q^{2\mfns \Delta}\rangle
\end{align}
where $| q^{2\mfns \Delta}\rangle$ denotes the two-sided TFD state 
associated to the operator  $q^{2\mfns\Delta}$ 
\begin{equation}
   | q^{2\mfns \Delta}\rangle \, \equiv\,\sum_n q^{2n\Delta}|n,n\rangle \,
\end{equation}
and $|0,0\rangle = |0\ra_L|0\ra_R$ denotes the factorized vacuum state in the tensor product  ${\cal H}_L \otimes {\cal H}_R$ of two DSSYK Hilbert spaces.
 The entangled state $| q^{2\mfns \Delta}\rangle$  can be created by acting with a suitable entangler $\cE_\Delta$ on the factorized vacuum state 
\begin{equation}
\label{entangler}
    \cE_\Delta|0,0\rangle = 
    | q^{2\mfns \Delta}\rangle, \qquad \qquad
\cE_\Delta = \frac{1}{(q^{4\Delta}  \yp_{{\!}_L} \yp_{{\!}_R}; q^2)_\infty\!\!}\;\; 
\end{equation}
with $\yp_L$ and $\yp_R$ the raising operators \eqref{raiselower} acting on ${\cal H}_L \otimes {\cal H}_R$. The alternative representation \eqref{gtwotwo} of the two-point function will turn out to be useful for writing an explicit expression for the out-of-time-ordered four-point function.

Inserting a complete set of energy eigenstates $|\theta_1,\theta_2\rangle$ into \eqref{gtwotwo} gives 
\begin{align}
G_\Delta(\beta_1,\beta_2) 
  & = \int_0^\pi \Bigl[\prod_{i=1,2} \!\frac{d\theta_i}{4\pi}  \sin^2\! \theta_i\, \mu(\theta)e^{-\beta_i E(\theta_i)}\Bigr] G_\Delta(\theta_1,\theta_2) \notag\\[2mm]
  &\hspace{-.7cm} G_\Delta(\theta_1,\theta_2)  = \langle \theta_1,\theta_2|  \cE_\Delta  |0,0\rangle = \la  \theta_1  | q^{2\mfns\Delta} |  \theta_2  \ra \label{melt}
  \end{align}
with $\mu(\theta)$ given in \eqref{mutheta}. Here we used that $\langle 0,0| \theta_1,\theta_2\rangle =1$.
The matrix element of $q^{2\mfns\Delta}$ between two energy eigenstates 
evaluates to
\be 
\label{gtwo}
G_\Delta(\theta_1,\theta_2)  = 
\frac{(q^{4\Delta}; q^2)_\infty\!}{(q^{2\Delta}\, e^{\pm i \theta_1\pm i \theta_2} ; q^2)_\infty\!}
\ee

Applying DSSYK-Schur triality to the two-point function gives useful insight on all three sides of the correspondence. First, we look for a  realization of the operator $D_\Delta$ that via the Schur-DSSYK dictionary maps to the two-point correlator
\bea
D_\Delta \quad \leftrightarrow \quad {\cal O}_{\Delta} {\cal O}_{\Delta} = q^{2\mfns\Delta} .
\eea

To gain some insight into what the two-point function represents on the ${\cal N}=2$ gauge theory side, let us replace the time evolution operators $e^{-\beta_i{\bf H}}$  by Wilson line operators $w_n = \chi_n({\bf H})$ with $\chi_n(x)$ the $n$-th Chebyshev polynomial. The two-point function associated to \eqref{gtwo} then takes the form
\begin{align}
{\!I\!\! I}_{n,m}(\Delta,q) 
& =  \int_0^\pi \Bigl[\prod_{i=1,2} \!\frac{d\theta_i}{4\pi}  \sin^2 \theta_i\Bigr] \;
{\!I\!\! I}(\theta_1,\theta_2,\Delta,q) \; w_{n}(\theta_1)w_{m}(\theta_2)
\label{dschur}
\end{align}
\be
\label{dintegrand}
{\!I\!\! I}(\theta_1,\theta_2,\Delta,q) = \,  \frac{(q^2,q^2)_\infty^2 (q^{4\Delta};q^2)_\infty (q^2 e^{\pm 2i\theta_1};q^2)_\infty(q^2 e^{\pm 2i\theta_2};q^2)_\infty}{  (q^{2\Delta} e^{\pm i\theta_1\pm i\theta_2};q^2)_\infty \strut}
\ee
We can identify the integral expression \eqref{dschur} as computing matrix elements of an operator defined by the integral kernel 
\begin{equation}
    K_\Delta(v_1,v_2) = \frac{(q^{4\Delta};q^2)_\infty}{(q^{2\Delta} v_1^\pm v_2^\pm;q^2)_\infty}
\end{equation}
We can act on this kernel by $\tH_a$:
\begin{align}
    &\tH_a K_\Delta(v_1,v_2) = \frac{q^{(a-1)/2}}{v_1^{-1}-v_1}\frac{(\fq^{4\Delta};\fq^2)_\infty}{(\fq^{2\Delta-1} v_1^\pm v_2^\pm;\fq^2)_\infty} \\[2mm] &\left[v_1^{a-1}(1-\fq^{2 \Delta-1}v_1 v_2)(1-\fq^{2 \Delta-1}v_1 v_2^{-1})-v_1^{1-a}(1-\fq^{2 \Delta-1}v_1^{-1} v_2)(1-\fq^{2 \Delta-1}v_1^{-1} v_2^{-1})\right]\ \nonumber
\end{align}
We recover the natural relation
\begin{equation}
    \tH_0 K_\Delta(v_1,v_2) = q^{-\frac12} K_{\Delta-\frac12}(v_1,v_2)
\end{equation}
Other relations are compatible with $K_\Delta$ formally representing 
$\tH_0^{-2{\Delta}}$.

We would like to give a physical interpretation to the expression \eqref{dschur}. The DSSYK matter chord operator $q^{2\mfns\Delta}$ formally looks like  a continuous negative power $T_0^{-2\Delta}$ of the 't Hooft operator $T_0$. However, this type of line operator does not exist in pure $SU(2)$ gauge theory. To build the operator $D_\Delta$ and reproduce the expression \eqref{dschur} for general continuous $\Delta$ from 4d gauge theory, we need to include more general building blocks in the form of a new kind of defect: 3d interfaces wrapping an $\bR^+\times \bC$ at some fixed angle $\varphi$ in the topological plane, ending at the Neumann boundary. The interface interpolates between two bulk sectors with Coulomb branch parameters $\theta_1$ and~$\theta_2$, as depicted in figure~5.

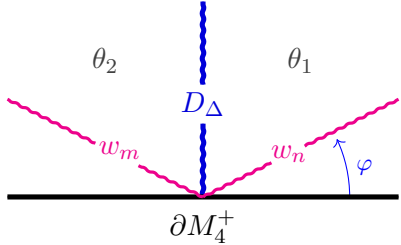
\begin{figure}[t]
\centering
\begin{tikzpicture}[xscale=1.3,yscale=1.3] 
\tikzset{
    partial ellipse/.style args={#1:#2:#3}{
        insert path={+ (#1:#3) arc (#1:#2:#3)}
    }
}   
\draw (1,1.4) node {$\textcolor{darkgray}{{\strut\ \theta_1\ }}$};
\draw (-1,1.4) node {$\textcolor{darkgray}{{\strut\ \theta_2\ }}$};
\draw[\darkblue,very thick,decorate, decoration={snake, segment length=1.5mm, amplitude=.1mm}] (-.01,0) -- (-.01,.75);
\draw[\darkblue] (0,.95) node {$D_\Delta$};
\draw[\darkblue,very thick,decorate, decoration={snake, segment length=1.5mm, amplitude=.1mm}] (-.01,1.15) -- (-.01,2);
\draw[\darkblue,very thick,decorate, decoration={snake, segment length=1.5mm, amplitude=.1mm}] (.01,0) -- (.01,.75);
\draw[\darkblue,very thick,decorate, decoration={snake, segment length=1.5mm, amplitude=.1mm}] (.01,1.15) -- (.01,2);
\draw[blue,->] (1.5,0) arc (0:25 :1.4) node[midway, right]  {{\footnotesize $\varphi$}};
\draw[very thick] (-2,0.01)--(2,0.01);
\draw[very thick] (-2,-0.01)--(2,-0.01);
\draw[magenta,very thick,decorate, decoration={snake, segment length=1.5mm, amplitude=.1mm}] (0,0) -- (-.66,.32);
\draw[magenta] (-.85,.48) node {$w_m$};
\draw[magenta,very thick,decorate, decoration={snake, segment length=1.5mm, amplitude=.1mm}] (-1.1,.55) -- (-2,1);
\draw[magenta,very thick,decorate, decoration={snake, segment length=1.5mm, amplitude=.1mm}] (0,0) -- (.7,.35);
\draw[magenta] (.88,.44) node {$w_n$};
\draw[magenta,very thick,decorate, decoration={snake, segment length=1.5mm, amplitude=.1mm}] (1.04,.52) -- (2,1);
\draw (0,-0.30) node {$\partial {M^+_4}$};
\end{tikzpicture}
\caption{A three point correlator involving two Wilson lines $w_n$ and $w_m$ and a domain wall operator $D_\Delta$. The domain wall interpolates between two sectors with different Coulomb branch parameters $\theta_1$ and $\theta_2$ and hosts a chiral multiplet with continuous $R$-charge $\Delta$.}
\end{figure}

We will now attempt to reverse engineer the gauge theory description of the interface operator $D_\Delta$  by deconstructing the explicit expression \eqref{dintegrand}.

The $\theta_i$-dependent numerator factors are appropriate to a Neumann boundary condition for the gauge fields on both sides of the interface. From the point of view of the gauge fields, the Neumann b.c. extend smoothly across the junction. 

The $\theta_i$-dependent denominator factors are identified with 3d chiral multiplets with a Neumann b.c. at the junction, transforming in a bi-fundamental representation of the two gauge groups. Here $\fq^\Delta$ plays the role of a fugacity for the $U(1)$ symmetry rotating the chiral multiplets. Physically, $\Delta = R + i \mathbb{R}$,
e.g. $\fq^\Delta = q^{R} \zeta$ for a phase $\zeta$ and an $R$-charge assignment $R$ for the chiral multiplets. 
These chiral multiplets come with a potential gauge anomaly 
at the junction, but that can be cancelled by a judicious choice of boundary Chern-Simons coupling. 

Finally, the $\Delta$-dependent numerator factor is naturally interpreted as the contribution of a 3d chiral multiplet with Dirichlet boundary conditions and appropriate R-charge and $U(1)$-charge $-2$. The charge assignment 
would be appropriate to add a boundary super-potential 
\begin{equation}
    W = \phi_{-2} \det \phi_1
\end{equation}
where we labelled the chirals by their $U(1)$ charge. Such a super-potential does not affect index calculations.

A good way to argue for such a super-potential is as folows. Notice that the choice $\Delta=0$ allows a superpotential deformation linear in $\phi_{-2}$. This forces $\phi_1$ to be invertible, which Higgses the interface to the trivial interface. Indeed, ${\cal O}_0$ is trivial. More generally, (half)integral negative $\Delta$ allows for position-dependent deformations enforcing co-dimension $2$ defects in the identity interface \cite{Gaiotto:2012xa}. We identify this defect with the 't Hooft line of charge $-2 \Delta \in \mathbb{N}$. For general values of $\Delta$ we can no longer reduce the configuration to a 1d defect and the interface will be a genuine 3d domain wall.\footnote{The combination of 5 hypermultiplets $\phi_{-2}$ and $\phi_1$ with superpotential $W = \phi_{-2} \det \phi_1 = \phi_{-2} (a d - b c)$
is a remarkable theory with many dual descriptions. In the 3d-3d correspondence setting, it is identified with an ideal octahedron built from $5$ tetrahedra.}

We can get more insight by applying the 3d-3d correspondence \cite{Dimofte:2011py,Dimofte:2013lba}. The 3d-3d correspondence has specific rules to characterize 3d manifolds triangulated by ideal tetrahedra which can represent interfaces for pure 4d $SU(2)$ gauge theory. Roughly, an interface corresponds to a 3d manifold with two $C$ boundaries. In practice, there is a rich set of rules to define the boundary of a triangulated three-manifold, which receive contributions both from the faces of the ideal tetrahedra and from small triangles cut out at the vertices of the tetrahedra. A ``$C$'' boundary component has a specific triangulation pattern which is described in \cite{Dimofte:2013lba}.

The 3d-3d correspondence associated an ideal tetrahedron to each chiral multiplet and glues the tetrahedra in a manner controlled by superpotential terms. Some faces of the resulting manifold may be still subject to further identifications to give the rest of the boundary the ``$C$'' patterns required for coupling to the 4d theory. The combination of five chiral multiplets with the super-potential term we proposed above as a 3d interface theory gives five ideal tetrahedra glued into an ideal octahedron. Four of the faces of the octahedron can be further glued together so that the remaining four faces and two of the cut vertices assemble into two ``$C$'' boundary components. 

The geometric details are a bit hard to visualize but they are a small variant of Figure 38 in \cite{Dimofte:2013lba}. That figure describes an octahedron glued into the geometry on the right side of Figure 37 in \cite{Dimofte:2013lba}. The geometry we find keeps the middle ``small'' cylinder connecting the two ``large'' two-pheres in that figure but replaces the ones going from a ``large'' two-sphere to itself
with two more connections between the two-spheres. The ``$C$'' components of the boundary consist respectively of the two left halves of the ``large'' two-spheres connected by a ``small'' cylinder, and of the two right halves connected by a ``small'' cylinder.

With some imagination, circles separating the two ``$C$'' components can be interpreted as representing two co-dimension 2 irregular defects which propagate in the 3d geometry connecting the irregular singularities in the two copies of $C$. As \cite{Dimofte:2013lba} does not define a dictionary for such irregular defects, this is a guess which should be supported by further work, which goes outside the scope of this paper. 
The middle ``small'' cylinder thus appear to stretch between the two irregular defects. Such small cylinders are known to represent Chern-Simons Wilson lines in infinite-dimensional representations with a second Casimir $c_2(\Delta)=  \Delta(\Delta-1)$.

We thus arrive at a very tentative class $S$ interpretation of the 3d interface corresponding to the insertion of  a matter chord $O_\Delta O_\Delta = \tH^{-2\Delta}_0$: it represents a configuration where the two irregular singularities in $C$ propagate from left to right, with an infinite-dimensional Wilson line ``rung'' stretched between them. This picture is very natural, considering that a 't Hooft line implements a similar configuration with a finite-dimensional Wilson line rung. 

 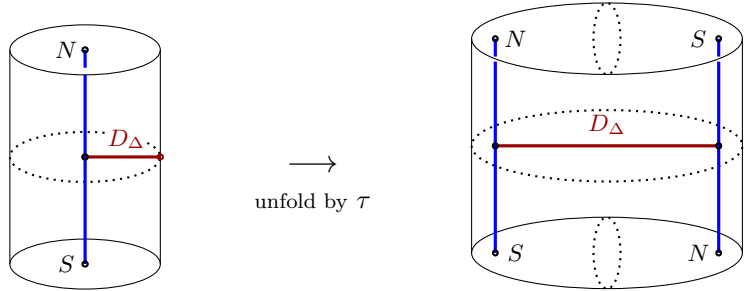
\begin{figure}[t]
 \centering
\begin{tikzpicture}[rotate=0,xscale=1,yscale=.95] 
\tikzset{
    partial ellipse/.style args={#1:#2:#3}{
        insert path={+ (#1:#3) arc (#1:#2:#3)}
    }
}   
\draw[black] (0,-1.5) [partial ellipse=-180:180:1cm and .35cm];
\draw[black] (0,1.5) [partial ellipse=-180:180:1cm and .35cm];
\draw[dotted,thick] (0,0) [partial ellipse=-285:35:1cm and .35cm];
\draw[blue,very thick] (0,-1.5) -- (0,1.15) ;
\draw[blue,very thick] (0,1.25) -- (0,1.5) ;
\draw[\darkred,very thick] (0,0) -- (1,0) ;
\draw (1,-1.5) -- (1,1.5);
\draw (-1,-1.5) -- (-1,1.5);
\draw[thick] (0,1.5) circle (.35mm);
\draw[thick] (0,-1.5) circle (.35mm);
\draw[thick,\darkred] (1,0) circle (.35mm);
\draw[thick,\darkred] (0,0) circle (.35mm);
\draw[thick] (0,0) circle (.35mm);
\draw (-0.25,-1.5)  node {\raisebox{0mm}{\footnotesize $S$}};
\draw (-0.25,1.5)  node {\raisebox{0mm}{\footnotesize $N$}};
\draw[\darkred] (0.55,.25) node {\mbox{\footnotesize $D_\Delta$}};
\end{tikzpicture}~~~~~~~~~\raisebox{1.3cm}{\LARGE ${\longrightarrow}\atop{\mbox{{\scriptsize unfold by}\small \  $\tau$}}$}~~~~~~~~~
\begin{tikzpicture}[rotate=0,xscale=.9,yscale=.95] 
\tikzset{
    partial ellipse/.style args={#1:#2:#3}{
        insert path={+ (#1:#3) arc (#1:#2:#3)}
    }
}   
\draw[black] (0,-1.5) [partial ellipse=-180:180:2cm and .5cm];
\draw[black] (0,1.5) [partial ellipse=-180:180:2cm and .5cm];
\draw[dotted,thick] (0,0) [partial ellipse=-180:180:2cm and .5cm];
\draw[dotted,thick] (0,1.5) [partial ellipse=-180:180:.2cm and .5cm];
\draw[dotted,thick] (0,-1.5) [partial ellipse=-180:180:.2cm and .5cm];
\draw[thick,\darkred] (-1.65,0) circle (.35mm);
\draw[thick,\darkred] (1.65,0) circle (.35mm);
\draw[\darkred,very thick] (-1.65,0) -- (1.65,0) ;
\draw[black] (2,-1.5) -- (2,1.5);
\draw[black] (-2,-1.5) -- (-2,1.5);
\draw[blue, very thick] (1.65,-1.5) -- (1.65,1.18) ;
\draw[blue, very thick] (1.65,1.25) -- (1.65,1.5) ;
\draw[blue,very thick] (-1.65,-1.5) -- (-1.65,1.18) ;
\draw[blue,very thick] (-1.65,1.25) -- (-1.65,1.5) ;
\draw[black,thick] (-1.65,-1.5) circle (.35mm);
\draw[black,thick] (1.65,0) circle (.35mm);
\draw[black,thick] (-1.65,0) circle (.35mm);
\draw[black,thick] (1.65,-1.5) circle (.35mm);
\draw[black,thick] (-1.65,1.5) circle (.35mm);
\draw[black,thick] (1.65,1.5) circle (.35mm);
\draw (-1.35,-1.5)  node {\raisebox{0mm}{\footnotesize $S$}};
\draw (-1.35,1.5)  node {\raisebox{0mm}{\footnotesize $N$}};
\draw (1.35,-1.5)  node {\raisebox{0mm}{\footnotesize $N$}};
\draw (1.35,1.5)  node {\raisebox{0mm}{\footnotesize $S$}};
\draw[\darkred] (0,.3) node {\mbox{\footnotesize $D_\Delta$}};
\end{tikzpicture}
\caption{In our proposed dictionary, the matter line operator $D_\Delta$ represents a $SL(2,C)$ Wilson line in the continuous series representation labeled by $\Delta$, placed as a "rung" connecting the irregular singularity with the cylindrical boundary.
Upon reversing the $\bZ_2$-identification $\tau$, the "rung" connects the irregular line defects inside $\widetilde{M}_3 = C \times I$. }
\end{figure}

The classical limit of this holonomy operator $D_\Delta$ can be formally represented  as an open $SL(2,\mathbb{C})$ Wilson line with end-points anchored to the irregular line defect 
\be
\label{ldelta}
D_\Delta = \bigl\la s'\bigr| {\rm P} \exp\Bigl(- \int_{\mbox{\tiny $S$}}^{\mbox{\tiny $N$}}\!\! A\Bigr) \bigl|s\bigr\ra 
\ee
evaluated between two states $|s\rangle$ and $|s'\rangle$ which lift to the representation the small sections $s$ and $s'$ available at the punctures. More precisely, these are highest weight states annihilated by the same lowering operators which annihilate $s$ and $s'$ respectively. For the spin 1/2 representation $\Delta=-1/2$ with Casimir $c_\Delta = 3/4$, the operator $D_\Delta$ reduces to the $SL(2,\mathbb{C})$ CS realization of the 't Hooft operator $\tH_0 = q^{-1/2-\mfns}$. For other irreducible finite-dimensional representations it reduces to a power of $\tH_0$. The non-diagonal matrix element \eqref{gtwo} of $D_\Delta$ between two energy eigenstates indicates that the line operator $D_\Delta$ does not commute with the monodromy operator $M$ around the punctures.

All of these considerations apply to the 3d interface itself, and need to be further refined to address the junction of the interface with Neumann boundary conditions and the role of the interface in producing an operator in the quantization of $SL(2,\mathbb{C})$ Chern-Simons theory on $C_\bR$. A natural guess is that the 3d picture above can be quotiented by $\tau$ to describe an infinite-dimensional rung stretched from the irregular singularity to the boundary of $C_\bR$, as for 't Hooft operators. 

We can now be more specific about the choice of infinite-dimensional representation in the rung. It should be a representation of $SL(2,\bC)$. In order for the Wilson line to end at the $SU(2)$ boundary naturally, it should admit a ``spherical'' $SU(2)$-invariant vector. Principal spherical representations for $SL(2,\bC)$ 
are labelled by the Casimir and are unitary on the $\Delta = \frac12 + i \bR$ line. 
In analogy with the finite-dimensional rungs, the endpoint on the irregular singularity should be an highest weight vector in a direction specified by $s$. 
These exist as unique distributional states in the principal spherical representations. We leave further tests of these ideas to future work.

\subsection{Out-of-time-order four point function}

\vspace{-1mm}

Finally, we make some preliminary comments about the potential Schur interpretation of the out-of-time ordered DSSYK four point function.

Let $V$ and $W$ denote local DSSYK operators with scale dimension $\Delta_V$ and $\Delta_W$. The corresponding out-of-time ordered four point function (OTOC) is defined as the crossed correator of the form
\be
\label{otoc}
{\rm OTOC}
= {\rm Tr}\Bigl[ W
e^{-\beta_4{\bf H}}
\, V e^{-\beta_3{\bf H}}\,  W e^{-\beta_2{\bf H}}\,V e^{-\beta_1{\bf H}}\Bigr] \, =\,  
\begin{tikzpicture}[xscale=-0.7,yscale=-.7, rotate=0, baseline={([yshift=-.07cm]current bounding box.center)}]
\draw[thick] (0,0) circle (1.5); 
\draw[thick,\darkblue] (1.5,.0) -- (0,.0); 
\draw[thick,\darkblue] (-1.5,.0) -- (0,.0);
\draw[thick,\darkred] (0,1.5) -- (0,0); 
\draw[thick,\darkred] (0,-1.5) -- (0,0);
\draw[fill,\darkred] (0,-1.5) circle (0.07); 
\draw[fill,\darkred] (0,1.5) circle (0.07);
\draw[fill,\darkblue] (-1.5,0) circle (0.07); 
\draw[fill,\darkblue] (1.5,0) circle (0.07);
\draw (1.3,1.3) node {\scriptsize $\beta_2$}; 
\draw (1.3,-1.3) node {\scriptsize $\beta_3$};
\draw (-1.3,1.3) node {\scriptsize $\beta_1$}; 
\draw (-1.3,-1.3) node {\scriptsize $\beta_4$};  
\draw (1.85,0) node {\small\textcolor{\darkblue}{$W$}}; 
\draw (0,1.8) node {\small \textcolor{\darkred}{$V$}};
\draw (-1.9,0) node {\small\textcolor{\darkblue}{$W$}}; 
\draw (0,-1.8) node {\small \textcolor{\darkred}{$V$}};
\end{tikzpicture}
\ee
The key feature that distinguishes the OTOC from the time-ordered four-point function is that the matter chords intersect and divide the thermal circle into four segments.

It is again useful to first look at the explicit expression for the correlator for guidance. The OTOC \eqref{otoc} can be computed in the double scaling limit by counting all the hamiltonian chords, with appropriate weight factors that take into account the presence of the intersecting pair of matter chords. The result can be conveniently expressed in terms of the entangler operator $\cE_\Delta$ introduced in \eqref{entangler} as follows \cite{Okuyama:2024yya,Okuyama:2024gsn}
\begin{align}
{\rm OTOC} 
& = \langle 0,0| 
e^{-\beta_3 {\bf H}_L-\beta_4 {\bf H}_R} 
 \cE_{\Delta_V}
\mfq^{(\mfns_L+\mfns_R)\Delta_W}
 \cE_{\Delta_V}^{-1} 
e^{-\beta_2 {\bf H}_L -\beta_1 {\bf H}_R}
| \mfq^{\mfns \Delta_V}\rangle
\\[2mm]
& =\int \Bigl[\prod_{i=1}^4  \frac{d\theta_i}{\pi}  \sin^2\! \theta_i\, \mu(\theta_i)\, 
\,e^{-\beta_i E(\theta_i)} 
\Bigr] \, R(\theta_1,\theta_2,\theta_3,\theta_4)\\[2.5mm]
& \hspace{-9mm}
R(\theta_1,\theta_2,\theta_3,\theta_4) = \langle \theta_3,\theta_4|\cE_{\Delta_V}
\mfq^{(\mfns_L+\mfns_R)\Delta_W}
 \cE_{\Delta_V}^{-1} |\theta_2,\theta_1\rangle \langle \theta_2| \mfq^{\mfns \Delta_V}|\theta_1\rangle
\end{align}
The crossing kernel $R(\theta_1,\theta_2,\theta_3,\theta_4)$ can be explicitly expressed in terms of the well-poised hypergeometric series ${}_8W_7$, which in turn is closely related to the $U_q(\mathfrak{su}(1,1))$ 6j-symbol  
\be
\biggl\{\begin{array} {ccc} \! {\Delta_V}\!& {s_1}\! & {s_2}\! \\ \!{\Delta_W}\!&  {s_3} \!& {s_4}\! \end{array} \biggr\}_q
\ee
with $s_i = \theta_i/\log q$. 

The appearance of the quantum 6j-symbol with $0<q<1$ in DSSYK correlators points to a correspondence with $SL(2,\mathbb{C})$ Chern-Simons theory. Indeed, one of the hallmark results in quantum CS theory is the identification of the q-deformed 6j-symbol with the expectation value of a tetrahedral configuration of Wilson lines \cite{Witten:1989rw}. The $SL(2,\mathbb{C})$ CS 6j-symbol for principal series representations is naturally identified with the super-conformal index of a specific 3d theory, identified in \cite{Dimofte:2013lba} from an ideal triangulation of the complement of a tetrahedral network of regular defects. See Figures 40 and 43 there. It would be interesting to probe the relation between the index or half-indices of that theory and the crossing kernel 
$R(\theta_1,\theta_2,\theta_3,\theta_4)$. Even better, one can attempt to express the OTOC in terms of an integral kernel which could be interpreted directly as the half-index of a 3d theory and mapped to a 3d manifold via the 3d-3d correspondence.

\section{Conclusion}

\vspace{-1mm}

We uncovered a precise correspondence between the double scaled SYK model and Schur correlators of 4d ${\cal N} =2$ $SU(2)$ gauge theory on a half-spacetime with Neumann boundary conditions.
 Applying the formalism of real Schur quantization, we have found that the correlator with $n$ fundamental Wilson lines on the boundary matches the $n$-th moment of the Hamiltonian ${\bf H}$ in the SYK model, while the square of the 't Hooft loop is identified with the product of two formal SYK operators ${\cal O}_{-1}$ with scale dimension $-1$. The new link between DSSYK and ${\cal N}=2$ SYM is part of larger set of dualities that include $SL(2,\mathbb{C})$ Chern-Simons theory with imaginary level compactified on a disk $C_\bR$ with a specific irregular singularity at the origin and a specific topological boundary condition.  This duality web is supported via a direct match between correlators and by the fact that the operator $*$-algebra $\cA_\fq$ in all three systems takes the form of the skein algebra Sk$_{q}(SL(2),C_{\bR})$ of line operators on $C_{\bR}$.  A direct by-product of our study is that it puts the link \cite{Verlinde:2024znh} between DSSYK and $SL(2,\mathbb{C})$ CS theory on firmer footing.

 \medskip

Our study leaves many open questions and directions to explore. We mention a few.
\vspace{-2mm}

\addtolength{\parskip}{-.8mm}

\begin{itemize}

\item{Our results hint that DSSYK sits inside the landscape of quantum systems that can be embedded in string theory. This opens a path to other potential dualities and physical realizations of DSSYK. Indeed, we could have anticipated our results via another duality sequence by realizing
$SU(2)$ SW theory as the decoupling limit of a string compactification on a Calabi-Yau  manifold with a local $A_1$ singularity \cite{Katz:1996fh}, as follows.  The mirror CY takes the form of a hyper surface $uv + y^2 = \lambda_{\rm SW}^2$ fibered over the SW curve $\Sigma_{\rm SW}$.  Via this realization, supersymmetric SW correlators are identified with topological string correlators on $\Sigma_{\rm SW}$. The $\mfq$-deformation leads to a refinement of the topological string  \cite{Aganagic:2011mi} that promotes the SW curve $y^2 = \lambda_{\rm SW}^2$ to a quantum curve $e^{\hat{y}} - \cos \hat{x} = v$ with $[\hat{x},\hat{y}] = i\hbar$. This SW quantum curve matches the eigenvalue equation of the DSSYK Hamiltonian \cite{Blommaert:2023wad}.}
\item{Can the complete DSSYK operator algebra, including general local operators ${\cal O}_\Delta$, be embedded as some natural BPS subalgebra of  ${\cal N}=2$ gauge theory? Our results appear to suggest that this might be possible. However, it would be somewhat surprising if this is true, because it would imply that a chaotic quantum many body system like SYK is dual to a supersymmetric quantum system. A possible resolution to this puzzle is that the DSSYK Hilbert space is BPS but that local operators ${\cal O}_\Delta$ are non-BPS, similar to the systems recently studied in \cite{Lin:2022rzw,Chen:2024oqv}. }

\item{What do our results teach us about a possible holographic interpretation of DSSYK? There exist a well-known formal relation between  $SL(2,\bC)$ CS theory and 3D de Sitter gravity \cite{Witten:1989ip,Castro:2011xb,Castro:2023bvo,Castro:2023dxp}. Both theories are exactly soluble and have correlators that can be exactly expressed as a sum over saddle-points. Moreover, the set of saddle-points of both theories overlap. Nonetheless, $SL(2,\bC)$ CS theory and 3D de Sitter gravity are not equivalent: the semiclassical expansion in each theory includes saddle points that do not naturally appear in the semiclassical expansion of the other. Hence DSSYK-Schur duality by itself does not imply that DSSYK is dual to de Sitter gravity. At the same time, our results do not exclude the possibility that a suitable class of SYK correlators can be given a dual interpretation in terms of de Sitter gravity. Indeed, it seems likely that comparing the two systems -- see where they match and where they might differ -- will reveal valuable new clues about de Sitter holography.}

\item{The disk $C_{\bR}$ in the construction can be replaced by other surfaces, giving rise to a variety of operator algebras and unitary representations. Can any of these be dual to variants of DSSYK?}
\end{itemize}

\medskip

\section*{Acknowledgements}
\vspace{-1mm}

We thank Monica Kang, Dongyeob Kim, Vladimir Narovlansky, Adel Rahman, Nathan Seiberg, Leonard Susskind, Joerg Teschner, Damiano Tieto, Erik Verlinde, Edward Witten, Jiuci Xu, and Mengyang Zhang for helpful discussions and comments. The research of H.V. is supported by NSF grant PHY-2209997.
The research of D.G. is supported by the NSERC Discovery Grant program and by the Perimeter Institute for Theoretical Physics. Research at Perimeter Institute is supported in part by
the Government of Canada through the Department of Innovation, Science and Economic
Development Canada and by the Province of Ontario through the Ministry of Colleges and
Universities.
\addtolength{\baselineskip}{-.6mm}

\bibliographystyle{ssg.bst}
\bibliography{Biblio.bib}

\end{document}